\documentclass[reprint,
superscriptaddress,
aps,
prb
]{revtex4-2}

\draft 

\usepackage{graphicx}

\usepackage{hyperref}

\hypersetup{
	colorlinks=true,
	linkcolor=blue,
	filecolor=magenta,      
	urlcolor=blue,
	citecolor=blue,
	pdfpagemode=FullScreen
}

\begin{document}



\title{Experimental and first-principles studies of superconductivity in topological nodal line semimetal SnTaS$_2$} 




\author{Soumen Ash}
\address{Department of Chemistry, Indian Institute of Technology Delhi, New Delhi 110016, India}
\affiliation{Institute of Nano Science and Technology, Mohali 140306, India}

\author{Moumita Naskar}
\affiliation{Department of Physics, Indian Institute of Technology Delhi, New Delhi 110016, India}

\author{Ravi Shankar P. N.}
\address{School of Advanced Materials; Chemistry and Physics of Materials Unit, Jawaharlal Nehru Centre for Advanced Scientific Research, Bangalore 560064, India}

\author{Nityasagar Jena}
\address{Laboratory for Chemistry of Novel Materials, Universit\'{e} de Mons, 7000 Mons, Belgium}


\author{A. Sundaresan}
\address{School of Advanced Materials; Chemistry and Physics of Materials Unit, Jawaharlal Nehru Centre for Advanced Scientific Research, Bangalore 560064, India}  


\author{Ashok Kumar Ganguli}
\email[E-mail: ]{ashok@chemistry.iitd.ac.in}
\affiliation{Department of Chemistry, Indian Institute of Technology Delhi, New Delhi 110016, India}
\affiliation{Department of Materials Science and Engineering, Indian Institute of Technology Delhi, New Delhi 110016, India}


\date{\today}

\begin{abstract}

We report a detailed study of superconductivity in polycrystalline SnTaS$_2$ using electrical transport, magnetization and heat capacity measurements. SnTaS$_2$ crystallizes in centrosymmetric hexagonal structure with space group $P6_3/mmc$. Electrical resistivity, magnetization and specific heat data suggest SnTaS$_2$ to be a weakly coupled, type-II superconductor with $T_c \approx$ 2.8 K. First-principles calculations show signature for nodal line topology in the electronic band structure, protected by the spatial-inversion and time-reversal symmetries, that strongly gapped out by the inclusion of spin-orbit coupling (SOC). Superconductivity in layered SnTaS$_2$ with nodal line topological state makes it a strong candidate to be considered for a 3D topological superconductor.

\end{abstract}


\maketitle 

\section{Introduction}



Novel quantum states like topological superconductors (TSCs) that host exotic excitations like Majorana fermions have stimulated intense research due to their fundamental physics and potential applications for quantum information technology \cite{Qi2011, Wang2018, Fu2008, Ando2015}. Topological superconductivity can be realized by two major approaches. First, by forming heterojunction between a topological material and a superconductor, in which topological surface states (TSSs) can host Cooper pairs and become superconducting through the proximity effect \cite{Fu2008, Akhmerov2009, Wang2012b, Hosur2011, Xu2014}. Second, by achieving superconductivity in a topological material or identifying the topological phase in a superconducting system \cite{Hor2010, Wray2010, Sasaki2011, Zhang2011, Zhu2013, Kirshenbaum2013, Shruti2015, He2016, Zhou2016}. Following the discovery of the topological insulators \cite{Chen2009, Zhang2009, Hasan2010, Hasan2011}, topological semimetals have recently attracted much attention due to their distinct symmetry protected nontrivial band characteristics \cite{Wang2012, Weng2015, Armitage2018, Soluyanov2015}. Based on the nature of the band crossing in the vicinity of the Fermi region, the topological semimetals are further classified into Dirac semimetals, Weyl semimetals and nodal line semimetals \cite{Chiu2016, Young2012, Lv2015, Schoop2016, Zhang2018, Weng2016}. In contrast to Weyl semimetals with essentially zero-dimensional (0D) bulk Fermi surface, nodal line semimetals possess extended band touching along a one-dimensional (1D) curve in the $k$-space and are expected to exhibit exotic topological physics \cite{Burkov2011, Lau2021}. Recent discovery of superconductivity in noncentrosymmetric topological nodal line semimetal PbTaSe$_2$ \cite{Ali2014, Bian2016, Zhang2016, Guan2016} has opened up a new avenue to explore the possibility of finding superconductivity and novel surface topology in related isoelectronic systems \cite{Chen2016, Chang2016, Wang2018b}. SnTaS$_2$ \cite{Chen2019, Feig2020, Chen2021} and PbTaS$_2$ \cite{Gao2020} with a centrosymmetric crystal structure are the latest additions to this family of intercalated 112-type transition metal dichalcogenide based topological nodal line superconductors.
%

The crystal structure and physical properties of SnTaS$_2$ were studied earlier in considerable detail \cite{Eppinga1977, Lee1990, Salvo1973, Eppinga1981, Dijkstra1989, Eppinga1976, Herber1975, Herber1976, Gentile1979, Herber1980, Gossard1974}. Albeit the superconducting critical temperature ($T_c \approx$ 2.5 - 2.95 K) \cite{Salvo1973, Eppinga1981, Dijkstra1989} and electronic structure \cite{Dijkstra1989, Guo1987, Blaha1991} of SnTaS$_2$ were known, a detailed investigation of its superconductivity and topological properties was not reported until recently \cite{Chen2019, Feig2020, Chen2021}. Although there are a few reports on the single crystal SnTaS$_2$, an extensive study on superconducting properties of the polycrystalline system is still required. 

It can thus be inferred that there is a research gap in terms of a comprehensive study on the superconducting properties of SnTaS$_2$ in polycrystalline form, and the aim of this work is to fill that gap. We have synthesized phase pure polycrystalline samples of SnTaS$_2$ via solid state reaction route. The structural, transport, magnetic and thermodynamic properties have been studied in detail to estimate the superconducting and normal-state parameters. We have also carried out first-principles calculations to investigate the electronic structure of the material. We found SnTaS$_2$ to be a weakly coupled, type-II superconductor with a topological nodal line state, which makes SnTaS$_2$ a potential candidate for 3D topological superconductor.

\section{Experimental and simulation details}

Polycrystalline samples of SnTaS$_2$ were synthesized by the solid state reaction of elemental Sn with prereacted TaS$_2$. TaS$_2$ precursor was prepared from the reaction of Ta powder (Alfa Aesar, 99.9\%) and S powder (Sigma Aldrich, 99.98\%) at 1123 K for 48 h in vacuum sealed condition. Next, Sn granules (Sigma Aldrich, 99.5+\%) and presynthesized TaS$_2$ powder were taken into stoichiometric ratio, ground thoroughly, vacuum sealed in a quartz ampoule, and heated at 1123 K for 48 h. The heat treated sample was then reground, pelletized, resealed in an evacuated quartz tube, and sintered at 1123 K for another 48 h. The final product is air-stable, dark grey in colour with a metallic lustre. The phase purity of the sample was verified by powder X-ray diffraction using a Bruker D8 Advance diffractometer with Cu-$K\alpha$ radiation. Structural refinement on room temperature powder X-ray diffraction data was carried out by the Rietveld method using the TOPAS software package \cite{Topas}. Elemental analysis was carried out by energy dispersive X-ray (EDX) spectroscopy using a Hitachi tabletop scanning electron microscope (SEM-EDX). A conventional four-probe technique was used for the transport property measurements, carried out in a physical properties measurement system (PPMS, Quantum Design). Field dependent and temperature dependent magnetization studies were conducted using a superconducting quantum interference device (SQUID, Quantum Design). The specific heat data were taken by the time relaxation technique using a physical property measurement system (PPMS, Quantum Design).

The electronic structure of SnTaS$_2$ was investigated by first-principles simulation using density functional theory (DFT) as implemented in the Vienna Ab initio Simulation Package (VASP) \cite{Kresse1993, Kresse1994, Kresse1996, Kresse1996b}. The generalized gradient approximation (GGA) in its Perdew-Burke-Ernzerhof (PBE) variant was adopted for the exchange-correlation functionals \cite{Kresse1994b, Kresse1999}. The experimental lattice parameters of SnTaS$_2$ were used in our DFT calculations to get the ground-state electron density with a plane-wave energy cutoff of 500 eV and the convergence criterion set at $10^{-6}$ eV. The Brillouin zone (BZ) was sampled using a 15 x 15 x 5 $\Gamma$-centered k mesh. The van der Waals (vdW) dispersion correction to the total energy expression was incorporated using the DFT-D3 method of Grimme \cite{Grimme2010, Grimme2011}, as implemented in VASP.


\section{Results and discussion}

\subsection{Structural properties}

\begin{figure}[!t]
\includegraphics[width= 1.0\columnwidth,angle=0,clip=true]{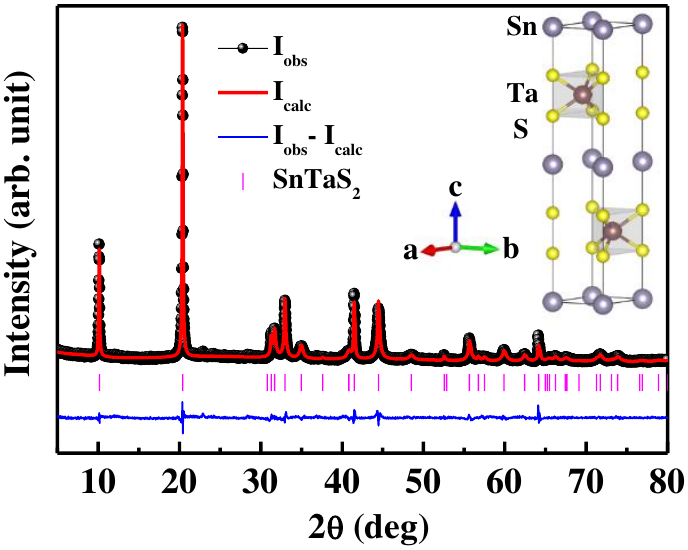}
\caption{(a) Rietveld refinement of the room temperature powder X-ray diffraction pattern of polycrystalline SnTaS$_2$. Vertical bars indicate the allowed Bragg's reflections for hexagonal $P6_3/mmc$ phase. Blue line indicates the difference between observed and fitted patterns. Inset: crystal structure of SnTaS$_2$.
}
\label{fig1}
\end{figure}

\begin{table}[!t]
\scriptsize\addtolength{\tabcolsep}{-1pt}
\caption{\label{table1}Refined structural parameters of SnTaS$_2$.}
\begin{ruledtabular}
\begin{tabular}{l c c c c c r}


\multicolumn{7}{l}{SnTaS$_2$} \\ 
\multicolumn{7}{l}{Space group: $P6_3/mmc$ (194)} \\
\multicolumn{7}{l}{{\it a = b} = 3.3006(2) \r{A} } \\
\multicolumn{7}{l}{{\it c} = 17.404(1) \r{A} }\\
\hline

Atom & Site & x & y & z & Occu. & B$_{eq}$ (\r{A}$^2$) \\
Sn & 2a & 0 & 0 & 0 & 1.04(1) & 0.5(2) \\
Ta & 2c & 1/3 & 2/3 & 1/4 & 1.00(1) & 1.1(2) \\
S & 4e & 0 & 0 & 0.1760(7) & 0.99(2) &  1.4(1) \\







\end{tabular}
\end{ruledtabular}
\end{table}

Rietveld refinement to the room temperature powder X-ray diffraction pattern of polycrystalline SnTaS$_2$ has been shown in Fig. \hyperref[fig1]{1}. Reliability factor of $R_{wp}$ = 7.20\% with a goodness of fit $\chi^2$ = 1.69\% demonstrate a reasonably good fitting. SnTaS$_2$ crystallizes in a centrosymmetric hexagonal structure with $P6_3/mmc$ space group. The refined lattice parameters [$a$ = 3.3006(2) \r{A} and $c$ = 17.404(1) \r{A}] for SnTaS$_2$ show adequate increase specifically in the $c$-lattice parameter compared to that for 2$H$-TaS$_2$ \cite{Meetsma1990}, suggests successful intercalation of Sn atoms in between the van der Waals (vdW) layers of TaS$_2$. Atomic ratio determined from the EDX spectra analysis is Sn$:$Ta$:$S = 1.02$:$1$:$1.82. A slight decrease in the lattice parameters compared to that reported earlier \cite{Eppinga1977, Lee1990} may be attributed to the small S-deficiency in the stoichiometry of our sample. Details of the refined structural parameters are given in Table \ref{table1}. 

The inset of Fig. \hyperref[fig1]{1} shows the crystal structure of SnTaS$_2$ consisting of alternating TaS$_2$ and Sn layers. In this structure, Ta occupies the trigonal-prismatic sites coordinated with $six$ S atoms, and Sn is in linear coordination with $two$ S atoms. The inversion symmetry in 2$H$-TaS$_2$ remains unaltered after the intercalation of Sn in the crystal structure.

\subsection{Transport properties}

\begin{figure}[!t]
\includegraphics[width= 1.0\columnwidth,angle=0,clip=true]{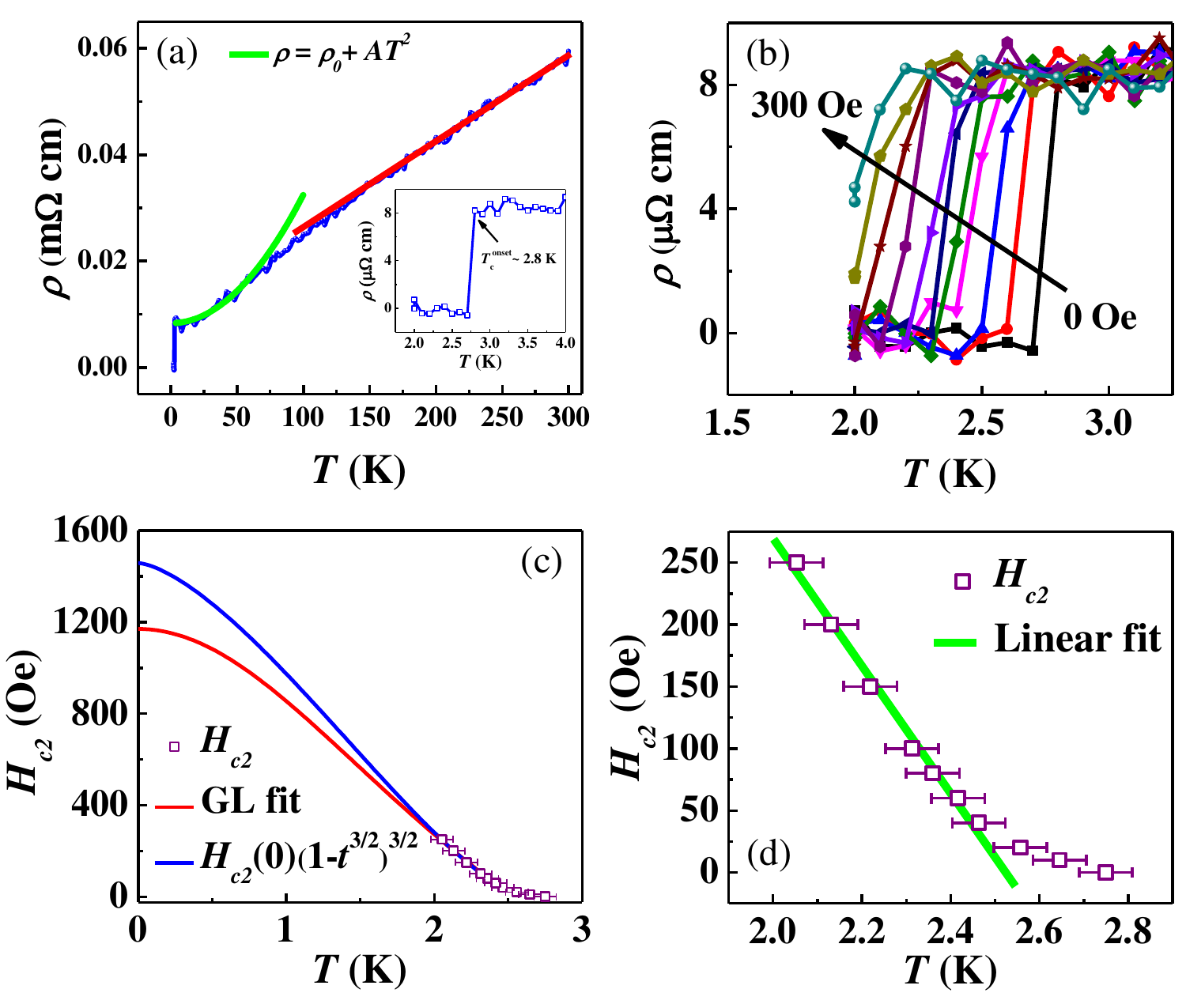}
\caption{(b) Temperature dependent resistivity of SnTaS$_2$. The solid red line is a guide to the linear behaviour of the resistivity in the high temperature range. The solid green line represents the $\rho = \rho_0+ AT^2$ fit. Inset: enlarged view of $\rho$ vs. $T$ data at low temperature showing the superconducting transition. (c) Superconducting transition from temperature dependent resistivity under various applied magnetic fields. (d) Temperature dependence of the upper critical field extracted from the resistivity plots. The solid red line represtents the Ginzburg-Landau fit and the solid blue line shows the fitting of the upper critical field data to $H_{c2}(T) = H_{c2}(0)(1 – t^{3/2})^{3/2}$ equation. Inset: linear fit (solid green line) to the temperature dependence of upper critical field data.
}
\label{fig2}
\end{figure}

Temperature dependence of resistivity at zero applied magnetic field has been studied for polycrystalline SnTaS$_2$ from 2 K to 300 K, as shown in Fig. \hyperref[fig2]{2}(a). The normal-state resistivity shows a metallic behaviour ($\frac{d\rho}{dT} > 0$) with a residual resistivity ratio RRR = $\rho_{300 K}/\rho_{3 K}$ = 6.8. The linear nature of the resistivity at high temperature has been shown by the red solid line, suggesting the dominance of electron-phonon scattering. Inset shows an enlarged view of low temperature resistivity data depicting a sharp superconducting drop with transition temperature $T_{c}^{onset} \approx$ 2.8 $\pm$ 0.05 K and transition width $\Delta T_c \approx$ 0.1 K. Temperature dependent resistivity data in the range of 3 – 60 K has been fitted with power law 

\begin{equation}\label{powerlaw}
\rho = \rho_0+ AT^2,
\end{equation} 

as shown by the green solid line in Fig. \hyperref[fig2]{2}(a). The $T^2$ dependence of resistivity in this temperature range suggests Fermi liquid behaviour of the system. From the fit, we found the residual resistivity $\rho_0$ = 8.34 $\pm$ 0.1 $\mu\Omega$ cm and coefficient $A$ = 0.0024 $\pm$ 0.0001 $\mu\Omega$ cm K$^{-2}$. The small value of $\rho_0$ indicates the highly conducting nature of the sample. 


Temperature dependent resistivity at varying magnetic fields has been studied in the temperature range of 2$–$4 K, as shown in Fig. \hyperref[fig2]{2}(b). A variable magnetic field of 0$–$300 Oe was applied perpendicular to the probe current. After finding the $T_c$ as the mid-point of the transition at different magnetic fields, the upper critical field values $H_{c2}$ have been plotted as a function of temperature shown in Fig. \hyperref[fig2]{2}(c). The high field temperature dependent $H_{c2}$ data has been fitted (red solid line) with generalized Ginzburg-Landau (GL) formula 
\begin{equation}\label{GL}
H_{c2}(T) = H_{c2}(0)(1 – t^2)/(1 + t^2), 
\end{equation}
where $t= T/T_c$ and $T_c$ is the transition temperature at zero magnetic field. The GL fit yields the upper critical field at $T$ = 0 K to be $\mu_{0}H_{c2}(0)$ = 1170 $\pm$ 6 Oe. Temperature dependent $H_{c2}$ data has an upward-like nature at the low temperature region and, therefore, considerably deviates from the GL fit. This upward feature possibly arises due to the multiband nature of the system \cite{Gurevich2003, Lei2012}. Similar upward-like behaviour of $H_{c2}(T)$ can also be seen for SnTaS$_2$ single crystal \cite{Chen2019, Feig2020} and other superconductors such as PbTaSe$_2$ \cite{Ali2014}, PbTaS$_2$ \cite{Gao2020}, Nb$_{0.18}$Re$_{0.82}$ \cite{Karki2011}, MgB$_2$ \cite{Takano2001}, La$_3$Se$_4$ \cite{Naskar2022} and borocarbides \cite{Lan2001, Rathnayaka1997}. Therefore, the $H_{c2}$ data has been further fitted (blue solid line) with
\begin{equation}\label{3/2}
H_{c2}(T) = H_{c2}(0)(1 – t^{3/2})^{3/2} 
\end{equation}
equation \cite{Micnas1990,Alexandrov1986,Alexandrov1993,Alexandrov2004}, where $t= T/T_c$, provides a better fit with $R^2$ = 0.99998. The upper critical field determined from this model is $\mu_{0}H_{c2}(0)$ = 1458 $\pm$ 2 Oe. This model has been used earlier for fitting the $H_{c2}(T)$ data of PbTaSe$_2$ \cite{Ali2014}, PbTaS$_2$ \cite{Gao2020}, Nb$_{0.18}$Re$_{0.82}$ \cite{Karki2011}, La$_3$Se$_4$ \cite{Naskar2022} and borocarbide superconductors \cite{Lan2001}. 



In case of a type-II superconductor, the breaking of Cooper pairs due to an externally applied magnetic field can be explained by two mechanisms: the orbital limiting effect and Pauli paramagnetic effect. In the orbital limiting effect, the field-induced kinetic energy of a Cooper pair exceeds the superconducting condensation energy, whereas, in the Pauli paramagnetic effect, the Zeeman splitting energy of the electrons exceeds the superconducting condensation energy resulting in Cooper pair breaking. For a single band BCS type superconductor, the orbital upper critical field can be determined from the Werthamer-Helfand-Hohenberg (WHH) formula 
\begin{equation}\label{WHH}
H_{c2}^{orb} = -AT_c\left(\frac{dH_{c2}}{dT}\right)_{T=T_c},
\end{equation}
where $A$ = 0.69 and 0.72 for dirty and clean limits, respectively \cite{Werthamer1966}. The linearly best fitted (green solid line) high field $H_{c2}(T)$ values, for the temperature range 2 K $< T <$ 2.5 K where $H_{c2}(T)$ shows a significant linear behaviour, are shown in Fig. \hyperref[fig2]{2}(d), and the slope is determined to be $\frac{dH_{c2}}{dT}$ = $-$ 510 $\pm$ 21 Oe/K. Hence taking $T_c$ = 2.75 K (the midpoint of the zero field resistivity transition), the orbital critical field is found to be $\mu_0H_{c2}^{orb}$ = 968 $\pm$ 57 Oe and 1010 $\pm$ 60 Oe for the dirty and clean limits, respectively. A similar approach was adapted for PbTaSe$_2$ \cite{Ali2014}, MgB$_2$ \cite{Takano2001}, La$_3$Se$_4$ \cite{Naskar2022}, RNi$_2$B$_2$C (R = Y, Lu) \cite{Rathnayaka1997} and Li$_2$(Pd$_{1-x}$Pt$_x$)$_3$B \cite{Peets2011} to determine the orbital critical field.

Further, if we consider the spin paramagnetic effect only, the Pauli limiting upper critical field \cite{Clogston1962, Chandrasekhar1962} can be determined from $H_P = \Delta/\sqrt2\mu_B$, where $\Delta$ = 1.76$k_BT_c$ for a weakly coupled BCS superconductor and $\mu_B$ is Bohr magneton. The Pauli paramagnetic limit for SnTaS$_2$ is found to be $\mu_0H_P$ = 5.1 $\pm$ 0.1 T. The characteristic Maki parameter, expressed as $\alpha = \sqrt2 H_{c2}^{orb}/H_P$ \cite{Maki1966}, provides a convenient measure for the relative strength of the orbital and Pauli pair breaking mechanisms. Using the value of $H_{c2}^{orb}$ and $H_P$, we found the value of $\alpha \approx$ 0.02. In our case, the value of the estimated upper critical field $H_{c2}(0)$ is comparable to the orbital critical field $H_{c2}^{orb}$, and in combination with a small value of the Maki parameter ($\alpha<<$ 1), suggests that the Pauli pair breaking is inconsiderable and the superconductivity in SnTaS$_2$ is limited by the orbital depairing effect.





\subsection{Magnetic properties}

\begin{figure}[!t]
\includegraphics[width= 1.0\columnwidth,height= 0.82\textheight,keepaspectratio,angle=0,clip=true]{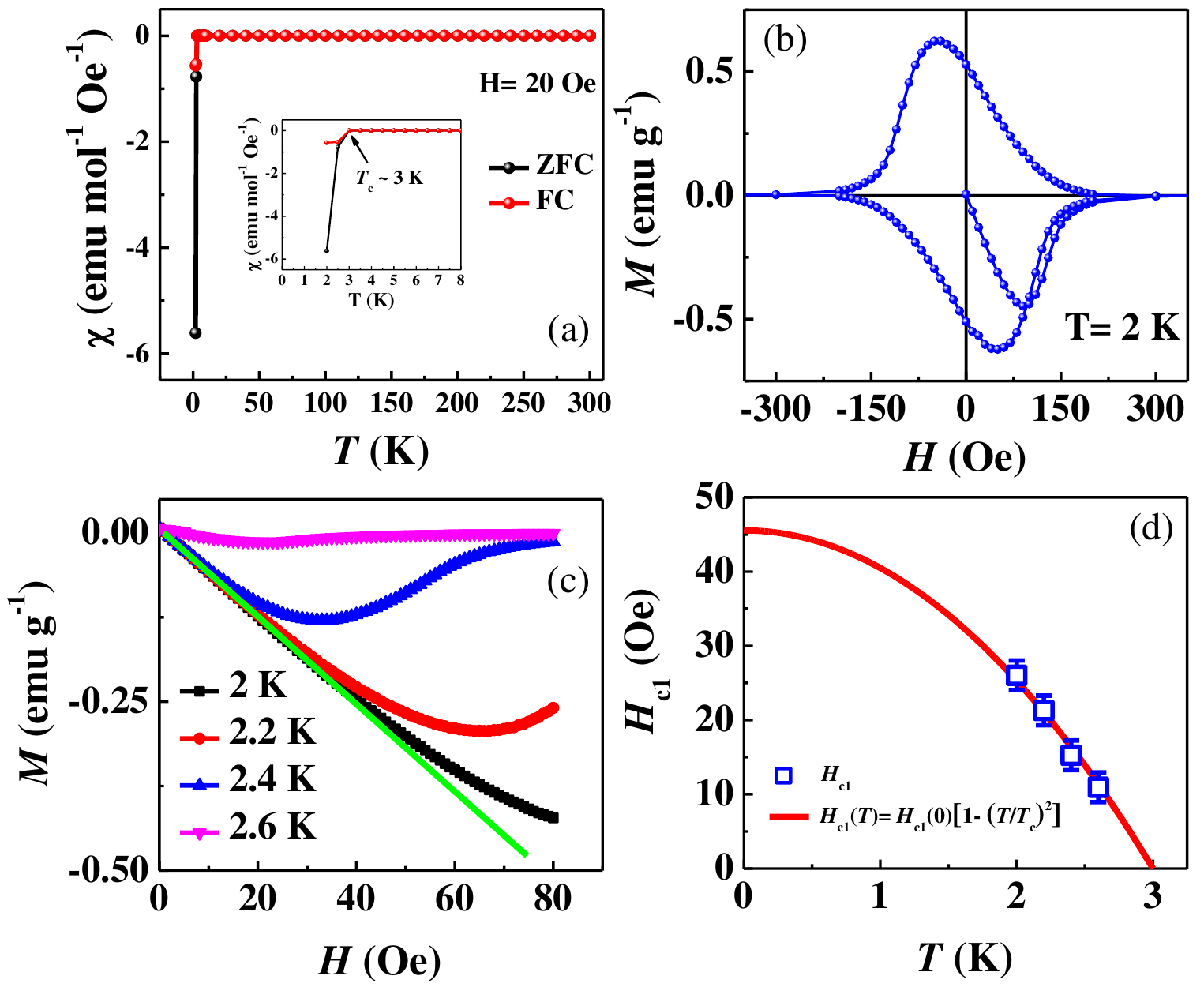}
\caption{(a) Temperature dependent zero field cooled (ZFC) and field cooled (FC) magnetic susceptibility with an applied magnetic field of 20 Oe. (b) Isothermal field dependent magnetization studied at 2 K. (c) Zero field cooled (ZFC) isothermal magnetization at different temperatures. The green solid line is the linear fit to the 2 K data in the low field range. (d) Lower critical field extracted from the isothermal magnetization curves. The solid red line represents the $H_{c1}(T) = H_{c1}(0)[1 – (T/T_c)^2]$ fit to the temperature dependence of the lower critical fields.
}
\label{fig3}
\end{figure}

Temperature dependent DC magnetic susceptibility of polycrystalline SnTaS$_2$ has been studied in the temperature range of 2 – 300 K under both zero field cooled (ZFC) and field cooled (FC) protocols with an applied magnetic field of 20 Oe as shown in Fig. \hyperref[fig3]{3}(a). Superconductivity below $\sim$ 3 K has been confirmed from the diamagnetic signals in the ZFC and FC curves, demonstrating the characteristic Meissner effect. The smaller diamagnetic signal for FC curve compared to ZFC data is attributed to vortex pinning in a type-II superconductor. Inset shows the enlarged view of the susceptibility data at low temperature.
Fig. \hyperref[fig3]{3}(b) depicts the isothermal field dependent magnetization of SnTaS$_2$ studied at 2 K. The magnetic hysteresis loop obtained suggests type-II superconductivity. The irreversible field estimated from the magnetization curve is $H_{irr}$ = 300 Oe at 2 K. Depinning of the vortices starts for $H > H_{irr}$.

Further, to determine the lower critical magnetic field of SnTaS$_2$, the field dependence of magnetization at various temperatures has been studied with a zero field cooling protocol. Fig. \hyperref[fig3]{3}(c) shows isothermal magnetization at different temperatures from 2$–$2.6 K, and the green solid line is the linear fit to the low field data. As we increase the applied magnetic field, the magnetization data starts to deviate from the linear fit, giving a lower critical field $H_{c1}$ for each isotherm. These lower critical field values are then fitted with the Ginzburg-Landau formula 
\begin{equation}\label{Hc1}
H_{c1}(T) = H_{c1}(0)(1 – t^2),
\end{equation}
where $t= T/T_c$ as shown by the red solid line in Fig. \hyperref[fig3]{3}(d). The zero temperature lower critical field is determined from the fit to be $\mu_0H_{c1}(0)$ = 45.5 $\pm$ 0.5 Oe.

Using $\mu_{0}H_{c2}(0)$ = 1458 $\pm$ 2 Oe for $T$= 0 K, the superconducting coherence length is estimated to be $\xi_{GL}(0) = [\Phi_0/{2\pi H_{c2}(0)}]^{1/2}$ = 47.5 $\pm$ 0.03 nm, where $\Phi_0 = h/2e$ ($h$ is the Planck's constant and $e$ is the charge of the electron). The Ginzburg-Landau superconducting penetration depth can be calculated from the formula $\mu_0H_{c1}(0) = (\Phi_0/4\pi\lambda_{GL}^{2})ln(\lambda_{GL}/\xi_{GL})$ and found to be $\lambda_{GL}(0) \approx$ 243 $\pm$ 1 nm. Further, the Ginzburg-Landau parameter to be determined as $\kappa = \lambda_{GL}(0)/\xi_{GL}(0)$ = 5.12 $\pm$ 0.02 ($\kappa > 1/\sqrt2$), confirms that SnTaS$_2$ is a type-II superconductor. The thermodynamic critical field of SnTaS$_2$ can be determined from the relation $H_{c1}(0)H_{c2}(0) = H_{c}^2(0)ln(\kappa)$ and found to be $\mu_0H_c(0)$ = 202 $\pm$ 1 Oe. The coherence length and the penetration depth values obtained for our polycrystalline sample lie within the corresponding values for the $H||c$-direction and $H||ab$ plane of the SnTaS$_2$ single crystal \cite{Chen2019}.


\subsection{Thermodynamic properties}

\begin{figure}[!t]
\includegraphics[width= 1.0\columnwidth,angle=0,clip=true]{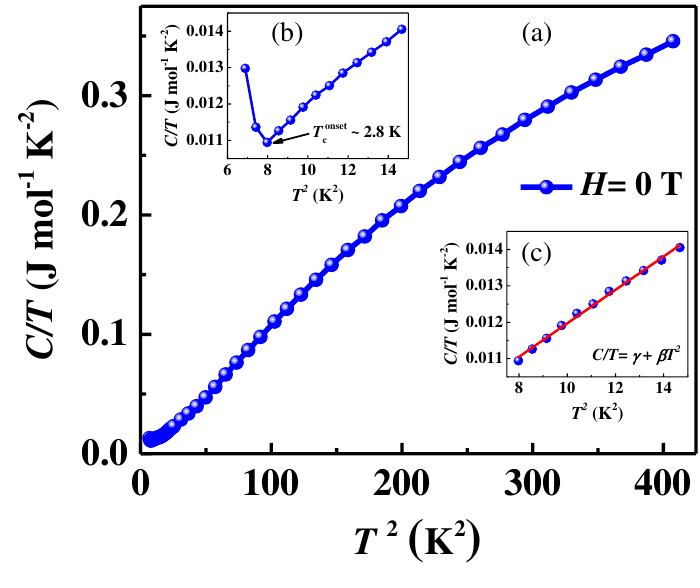}
\caption{Temperature dependent specific heat measured at $H$ = 0 T and represented as $C/T$ vs. $T^2$. (b) $C/T$ vs. $T^2$ data at $H$ = 0 T showing the initial of the bulk superconducting jump at $T_c^{onset}\sim$ 2.8 K. (c) Low temperature $C/T$ vs. $T^2$ data fitted with $C/T = \gamma + \beta T^2$ equation.
}
\label{fig4}
\end{figure}

\begin{table*}[!t]
\scriptsize\addtolength{\tabcolsep}{-5pt}
\caption{\label{table2}Normal-state and superconducting parameters of polycrystalline SnTaS$_2$ and other related isoelectronic systems.}
\centering
\begin{ruledtabular}
\begin{tabular}{l c c c c c c r}

Parameter&Unit&PbTaSe$_2$\cite{Ali2014}&PbTaSe$_2$\cite{Zhang2016}&PbTaS$_2$\cite{Gao2020}&SnTaS$_2$\cite{Chen2019}&SnTaS$_2$\cite{Feig2020}&SnTaS$_2$$^\dagger$\\
\hline
Material& &Polycrystalline&Single crystal&Single crystal&Single crystal&Single crystal&Polycrystalline\\
Space group& &$P\bar{6}m2$&$P\bar{6}m2$&$P6_3/mmc$&$P6_3/mmc$&$P6_3/mmc$&$P6_3/mmc$\\
$T_c$&K&3.72&3.8&2.6&3&2.9&2.8\\
$\mu_0H_{c1}(0)$&mT&7.5&9&3.4$^a$,4.1$^b$&5.4$^a$,8.2$^b$&\textcolor{blue}{$ $}&4.5\\
$\mu_0H_{c2}(0)$&T&1.47&1.25$^a$,0.32$^b$&0.32$^a$,0.05$^b$&0.3$^a$,0.02$^b$&0.17$^a$,0.03$^b$&0.15\\
$\mu_0H_P$&T&6.8&6.9&4.84&5.5&\textcolor{blue}{$ $}&5.09\\
$\xi_{GL}(0)$&nm&15&16.2&85.1$^a$,12.2$^b$&127$^a$,8.5$^b$&110$^a$,17$^b$&47.5\\
$\lambda_{GL}(0)$&nm&248&\textcolor{blue}{$ $}&127.3$^a$,889.7$^b$&64.4$^a$,962.4$^b$&\textcolor{blue}{$ $}&243\\
$\kappa$& &17&\textcolor{blue}{$ $}&10.5$^a$,1.5$^b$&7.6$^a$,5.1$^b$&3$^a$&5.12\\

$\gamma$&mJ mol$^{-1}$K$^{-2}$&6.9&6.01&6.07&4.45&5.4&7.37\\
$\Theta_D$&K&112&161&120&154.4&229&257\\
$\lambda_{ep}$& &0.74&\textcolor{blue}{$ $}&0.69&0.66&\textcolor{blue}{$ $}&0.57\\

\hline
$^\dagger:$ this work\\
$^a: H//ab$\\
$^b: H//c$\\

\end{tabular}
\end{ruledtabular}
\end{table*}

To further investigate the normal-state and superconductivity in SnTaS$_2$, temperature dependent specific heat measurement has been carried out at zero magnetic field, presented as $C/T$ vs. $T^2$ in Fig. \hyperref[fig4]{4}(a). Due to the limitation of our instrument, we could not go below 2.6 K to capture the entire specific heat anomaly due to the superconducting transition [Fig. \hyperref[fig4]{4}(b)]. However, the onset of the anomaly can certainly be observed at $T_c^{onset}\approx$ 2.8 K, confirming bulk superconductivity in the system.
%
The low temperature specific heat data, just above the superconducting transition, allows us to fit the $C/T$ vs. $T^2$ curve to the relation 
\begin{equation}\label{CT}
C/T = \gamma + \beta T^2,
\end{equation}
as shown by the red solid line in Fig. \hyperref[fig4]{4}(c). From the fit, we can determine the normal state Sommerfeld coefficient, $\gamma$ = 7.37 $\pm$ 0.11 mJ mol$^{-1}$ K$^{-2}$ and the phonon-specific coefficient, $\beta$ = 0.46 $\pm$ 0.01 mJ mol$^{-1}$ K$^{-4}$. 


From the relation
\begin{equation}\label{Theta}
\Theta_D = \left[\frac{12}{5\beta}\pi^4nN_Ak_B\right]^{1/3},
\end{equation}
where n = 4 for SnTaS$_2$, $N_A$ is the Avogadro number, $k_B$ is the Boltzmann’s constant, and using the value of $\beta$, the characteristic Debye temperature is estimated to be $\Theta_D$ = 257 $\pm$ 2 K. The electron-phonon coupling constant $\lambda_{ep}$ can be determined using McMillan’s formula \cite{Mcmillan1968}, 
\begin{equation}\label{lambda}
\lambda_{ep} = \frac{1.04+\mu^{*} ln(\Theta_D/1.45T_c)} {(1 - 0.62 \mu^*) ln(\Theta_D/1.45T_c)-1.04},
\end{equation}
where $\mu^*$ is the Coulomb pseudopotential. Taking $\mu^*$ = 0.13 for transition metal based superconductors, the value of $\lambda_{ep}$ is found to be 0.57 $\pm$ 0.01 which is less than 1, the minimum value for strong coupling, suggests SnTaS$_2$ is a weakly coupled superconductor. Using these results one can derive the noninteracting density of states at the Fermi level from $N(E_F) = 3\gamma/[\pi^2k_B^2(1+ \lambda_{ep})]$, giving $N(E_F) \approx$ 2.0 $\pm$ 0.04 states eV$^{-1}$ f.u.$^{-1}$ (f.u.: formula unit). 

The Kadowaki-Woods ratio (KWR) = $A/\gamma^2$, where $A$ is the coefficient of the quadratic term in the temperature dependent resistivity (Eq. \hyperref[powerlaw]{1}) and $\gamma$ is the Sommerfeld coefficient, is considered as a measure of the degree of electron correlations \cite{Kadowaki1986}. In heavy fermionic systems where electron-electron correlation is significant KWR approaches a nearly universal value $a_0$ = 1 $\times$ 10$^{-5}$ $\mu\Omega$ cm mJ$^{-2}$ mol$^2$ K$^2$ \cite{Kadowaki1986, Jacko2009}. Using $A$ = 0.0024 $\mu\Omega$ cm K$^{-2}$ and $\gamma$ = 7.37 mJ mol$^{-1}$ K$^{-2}$, in our case, we obtained KWR $\approx$ 4.42$a_0$ which suggests SnTaS$_2$ is a strongly correlated electron system.


Table \hyperref[table2]{II} summarizes the normal-state and superconducting parameters of the polycrystalline SnTaS$_2$ obtained from our studies, and compares them with other isoelectronic systems. A polycrystalline material is comprised of several individual grains or crystallites. These grains can be considered as single crystals with their correspondent crystallographic planes misaligned with respect to each other at random degrees. Therefore, for an isotropic polycrystalline solid, the measured physical properties get averaged over all the crystallographic directions. SnTaS$_2$ is an anisotropic superconductor, where the upper critical field measured parallel to the $ab$-plane [$H_{c2}^{ab}(0)$ $\approx$ 0.17 $-$ 0.3 T] is much larger than that measured parallel to the $c$-direction [$H_{c2}^{c}(0)$ $\approx$ 0.02 $-$ 0.03 T] \cite{Chen2019, Feig2020}. In our case, the effect of anisotropy get distributed over all the crystallographic directions for the polycrystalline sample, and hence the upper critical field is found to be $\approx$ 0.15 T. Similarly, the estimated values of the superconducting coherence length [$\xi_{GL}(0)$] and the penetration depth [$\lambda_{GL}(0)$] are found intermediate to their corresponding direction dependent values obtained for the single crystal system \cite{Chen2019, Feig2020}. Furthermore, the superconducting $T_c$ found for the polycrystalline sample ($\approx$ 2.8 K) is slightly lower than that of the reported single crystal ($\approx$ 2.9 $-$ 3 K) \cite{Chen2019, Feig2020, Singh2022}. Also, the irreversibility in the isothermal magnetization curve obtained at $T$ = 2 K [Fig. \hyperref[fig3]{3}(b)] indicates a substantial flux pinning behavior in the polycrystalline system, which is different from the behavior reported for the single crystal system \cite{Feig2020, Singh2022}. The marginal decrease in $T_c$ and considerable flux pinning behavior may be attributed to grain boundaries and possible defects in the polycrystalline sample. Additionally, the estimated Sommerfeld coefficient ($\gamma$) and the Debye temperature ($\Theta_D$) of the polycrystalline system are found to be higher than that reported for the single crystals \cite{Chen2019, Feig2020}. Although there are no major differences in the overall physical properties of the systems, the superconducting and normal-state parameters are found to be dependent on the single crystal and the polycrystalline nature of the samples.

\subsection{Theoretical calculations}

\begin{figure}[!t]
\includegraphics[width= 1.0\columnwidth,angle=0,clip=true]{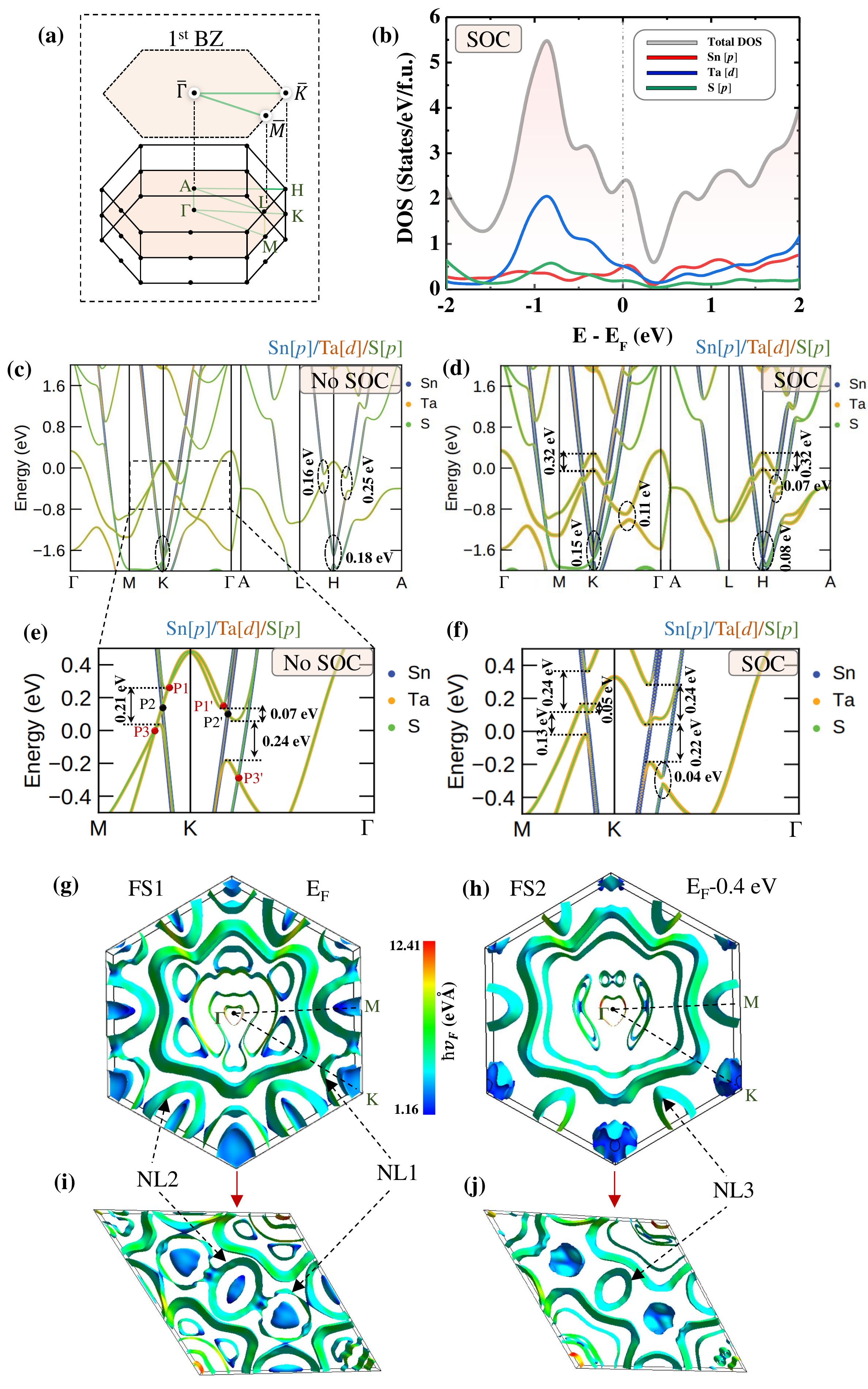}
\caption{(a) The schematic represents the bulk 3D Brillouin zone of SnTaS$_2$ and the corresponding 2D projection with high symmetry points. (b) Total and orbital-projected DOS with SOC. Electronic band structure of SnTaS$_2$ (c) without and (d) with SOC effects. The corresponding atomic and orbital components of Sn($p$), Ta($d$) and S($p$) orbitals are depicted in different colours such as blue, orange and green, respectively. Enlarged view of the low-energy electronic structure in the vicinity of Fermi energy ($E_F \pm 0.5$ eV) along the $M - K$ and $K - \Gamma$ high-symmetry lines (e) without and (f) with SOC effects taken into account. (g) Fermi surface (FS1) at $E_F(= 0)$ and (h) Fermi surface (FS2) at $E_F - 0.4$ eV without the SOC. (i)-(j) Slices of Fermi surfaces at $E_F$ and $E_F - 0.4$ eV, respectively. Slices are plotted in the primitive cell in order to show the closed orbits. The nodal lines (NL1, NL2 and NL3) are indicated by the arrows.
}
\label{fig5}
\end{figure}


Fig. \hyperref[fig5]{5}(a) shows the three-dimensional (3D) and the projected two-dimensional (2D) Brillouin zone (BZ) with $\Gamma$, $K$, and $M$ points serving as the high symmetry points for the $k_z$ = 0 plane and $A, H$ and $L$ points for the $k_z$ = $\pi$ plane. Fig. \hyperref[fig5]{5}(b) represents the total and orbital-projected density of states (DOS) with spin-orbit coupling (SOC). The characteristic Fermi level lies on a local maxima of the DOS, resulting nonzero states at Fermi energy. The Fermi energy level is largely dominated by the contribution from Ta and Sn, giving the total number of electronic states $N({\rm E_F})_{band}$ = 2.37 states eV$^{-1}$ f.u.$^{-1}$. This value is comparable to the DOS estimated from the specific heat data. 
Within the framework of the lowest order Sommerfeld expansion of electronic specific heat, the theoretical value of the Sommerfeld coefficient can be obtained from the relation $\gamma_{band}$ = $\frac{\pi^2}{3}k_B^2N({\rm E_F})_{band}$. We found $\gamma_{band}$ = 5.6 mJ mol$^{-1}$ K$^{-2}$, which can be correlated to the experimental value $\gamma$ = 7.37 mJ mol$^{-1}$ K$^{-2}$ by the relation $\gamma$ = $\gamma_{band}(1 + \lambda_{ep})$, giving $\lambda_{ep} \approx$ 0.32. The value of $\lambda_{ep}$ ($<$ 1) estimated from the theoretical data further confirms weak electron-phonon coupling in SnTaS$_2$, which is also inferred from our specific heat study.

Fig. \hyperref[fig5]{5}(c)-(d) shows the orbital projected band structure of SnTaS$_2$ without and with the inclusion of spin-orbit coupling (SOC) effects. Several electronlike and holelike band crossings can be observed in the Fermi level, consistent with the multiband characteristic \cite{Gurevich2003}, possible origin for the upward feature of the $H_{c2}(T)$ in our sample. Near the Fermi energy ($E_F \pm 0.4$ eV), topological nodal-line (NL) semimetallic character have appeared in SnTaS$_2$ by the crossing of several bands to form three nodal lines in the momentum space ($k_z$ = 0 plane).  In SnTaS$_2$, the topological NL states are majorly contributed by the Ta ($5d$) and Sn ($5p$) orbitals and protected by the spatial-inversion symmetry and time-reversal symmetry around the $K$-point without explicit consideration of the SOC effects, as shown in Fig. \hyperref[fig5]{5}(c), (e). The band topology of SnTaS$_2$ was further examined in Fig. \hyperref[fig5]{5}(e)-(f) in an energy range of $E_F \pm 0.5$ eV along the high symmetry line, $M-K-\Gamma$. Four crossing points (CP)- $P1, P1', P2, P2'$ are found to lie above the Fermi level, while $P3$ and $P3'$ nodes are buried below the $E_F$. The CP- $P2, P2'$ exhibit linearity with \textbf{\textit{k}}, while the CP- $P1, P1', P3, P3'$ shows a quadratic dependence on \textbf{\textit{k}}. Under SOC, these nodal lines are strongly gapped out in energy scales $\sim 40-240$ meV, as shown in Fig. \hyperref[fig5]{5}(f), along the $M-K$ and $K-\Gamma$ high-symmetry lines due to the difference in the SOC strength of Ta and Sn. 

The formation of NLs due to the band-crossings in the momentum space, in Fig. \hyperref[fig5]{5}(g)-(h), has been shown by the constant energy surfaces at the $E_F$ (FS1) and the same at the $E_F-0.4$ eV (FS2), respectively, with the velocity scaling. The primitive cell cuts of the corresponding Fermi surfaces are presented in Fig. \hyperref[fig5]{5}(i) and (j), respectively. The nodal lines NL1, NL2 and NL3 formed by the crossing points $P1(P1')$, $P2(P2')$ and $P3(P3')$, respectively, are indicated by the arrows shown in Fig. \hyperref[fig5]{5}(g)-(j). Fig. \hyperref[fig5]{5}(i)-(j) show the extremal closed orbits at FS1 and FS2, demonstrating the nodal line topology in SnTaS$_2$.



\section{Conclusions}

In conclusion, we present a comprehensive study on the superconducting and normal-state properties of polycrystalline SnTaS$_2$ using electrical transport, magnetization, and specific heat measurements. We report type-II bulk superconductivity ($T_c \approx$ 2.8 K) in polycrystalline SnTaS$_2$. Due to the polycrystalline nature of the sample, the measured physical properties are averaged over all the crystallographic directions. The estimated upper critical field, coherence length, and penetration depth are in between their corresponding direction dependent values. Specific heat data shows that SnTaS$_2$ is a weakly coupled superconductor. The KWR value suggests a strong electron-electron correlation in SnTaS$_2$. The topological nature of its electronic structure has been studied using density functional theory. In the vicinity of the Fermi level, topological nodal lines appeared due to the crossings of bands majorly contributed from Ta($5d$) and Sn($5p$) orbitals. The nodal lines get strongly gapped out under the influence of SOC. Superconductivity in this layered system which satisfies the requirements of a topological nodal line semimetal, offers the exciting possibility of realizing 3D topological superconductor and Majorana fermions in SnTaS$_2$.




%
%

%

\begin{acknowledgments}
We acknowledge CRF at IIT Delhi for the SQUID facility. AS thanks the ICMS and SSL at JNCASR, Bangalore for providing experimental facilities. SA acknowledges INST, Mohali for fellowship. MN acknowledges IIT Delhi for fellowship. AKG thanks SERB-DST, India for financial support (project sanction no.: EMR/2016/000156). 


\end{acknowledgments}



\begin{thebibliography}{87}%
\makeatletter
\providecommand \@ifxundefined [1]{%
 \@ifx{#1\undefined}
}%
\providecommand \@ifnum [1]{%
 \ifnum #1\expandafter \@firstoftwo
 \else \expandafter \@secondoftwo
 \fi
}%
\providecommand \@ifx [1]{%
 \ifx #1\expandafter \@firstoftwo
 \else \expandafter \@secondoftwo
 \fi
}%
\providecommand \natexlab [1]{#1}%
\providecommand \enquote  [1]{``#1''}%
\providecommand \bibnamefont  [1]{#1}%
\providecommand \bibfnamefont [1]{#1}%
\providecommand \citenamefont [1]{#1}%
\providecommand \href@noop [0]{\@secondoftwo}%
\providecommand \href [0]{\begingroup \@sanitize@url \@href}%
\providecommand \@href[1]{\@@startlink{#1}\@@href}%
\providecommand \@@href[1]{\endgroup#1\@@endlink}%
\providecommand \@sanitize@url [0]{\catcode `\\12\catcode `\$12\catcode
  `\&12\catcode `\#12\catcode `\^12\catcode `\_12\catcode `\%12\relax}%
\providecommand \@@startlink[1]{}%
\providecommand \@@endlink[0]{}%
\providecommand \url  [0]{\begingroup\@sanitize@url \@url }%
\providecommand \@url [1]{\endgroup\@href {#1}{\urlprefix }}%
\providecommand \urlprefix  [0]{URL }%
\providecommand \Eprint [0]{\href }%
\providecommand \doibase [0]{https://doi.org/}%
\providecommand \selectlanguage [0]{\@gobble}%
\providecommand \bibinfo  [0]{\@secondoftwo}%
\providecommand \bibfield  [0]{\@secondoftwo}%
\providecommand \translation [1]{[#1]}%
\providecommand \BibitemOpen [0]{}%
\providecommand \bibitemStop [0]{}%
\providecommand \bibitemNoStop [0]{.\EOS\space}%
\providecommand \EOS [0]{\spacefactor3000\relax}%
\providecommand \BibitemShut  [1]{\csname bibitem#1\endcsname}%
\let\auto@bib@innerbib\@empty
\bibitem [{\citenamefont {Qi}\ and\ \citenamefont {Zhang}(2011)}]{Qi2011}%
  \BibitemOpen
  \bibfield  {author} {\bibinfo {author} {\bibfnamefont {X.-L.}\ \bibnamefont
  {Qi}}\ and\ \bibinfo {author} {\bibfnamefont {S.-C.}\ \bibnamefont {Zhang}},\
  }\href {https://doi.org/10.1103/RevModPhys.83.1057} {\bibfield  {journal}
  {\bibinfo  {journal} {Rev. Mod. Phys.}\ }\textbf {\bibinfo {volume} {83}},\
  \bibinfo {pages} {1057} (\bibinfo {year} {2011})}\BibitemShut {NoStop}%
\bibitem [{\citenamefont {Wang}(2018)}]{Wang2018}%
  \BibitemOpen
  \bibfield  {author} {\bibinfo {author} {\bibfnamefont {J.}~\bibnamefont
  {Wang}},\ }\href {https://doi.org/10.1093/nsr/nwy155} {\bibfield  {journal}
  {\bibinfo  {journal} {Natl. Sci. Rev.}\ }\textbf {\bibinfo {volume} {6}},\
  \bibinfo {pages} {199} (\bibinfo {year} {2018})}\BibitemShut {NoStop}%
\bibitem [{\citenamefont {Fu}\ and\ \citenamefont {Kane}(2008)}]{Fu2008}%
  \BibitemOpen
  \bibfield  {author} {\bibinfo {author} {\bibfnamefont {L.}~\bibnamefont
  {Fu}}\ and\ \bibinfo {author} {\bibfnamefont {C.~L.}\ \bibnamefont {Kane}},\
  }\href {https://doi.org/10.1103/PhysRevLett.100.096407} {\bibfield  {journal}
  {\bibinfo  {journal} {Phys. Rev. Lett.}\ }\textbf {\bibinfo {volume} {100}},\
  \bibinfo {pages} {096407} (\bibinfo {year} {2008})}\BibitemShut {NoStop}%
\bibitem [{\citenamefont {Ando}\ and\ \citenamefont {Fu}(2015)}]{Ando2015}%
  \BibitemOpen
  \bibfield  {author} {\bibinfo {author} {\bibfnamefont {Y.}~\bibnamefont
  {Ando}}\ and\ \bibinfo {author} {\bibfnamefont {L.}~\bibnamefont {Fu}},\
  }\href {https://doi.org/10.1146/annurev-conmatphys-031214-014501} {\bibfield
  {journal} {\bibinfo  {journal} {Annu. Rev. Condens. Matter Phys.}\ }\textbf
  {\bibinfo {volume} {6}},\ \bibinfo {pages} {361} (\bibinfo {year}
  {2015})}\BibitemShut {NoStop}%
\bibitem [{\citenamefont {Akhmerov}\ \emph {et~al.}(2009)\citenamefont
  {Akhmerov}, \citenamefont {Nilsson},\ and\ \citenamefont
  {Beenakker}}]{Akhmerov2009}%
  \BibitemOpen
  \bibfield  {author} {\bibinfo {author} {\bibfnamefont {A.~R.}\ \bibnamefont
  {Akhmerov}}, \bibinfo {author} {\bibfnamefont {J.}~\bibnamefont {Nilsson}},\
  and\ \bibinfo {author} {\bibfnamefont {C.~W.~J.}\ \bibnamefont {Beenakker}},\
  }\href {https://doi.org/10.1103/PhysRevLett.102.216404} {\bibfield  {journal}
  {\bibinfo  {journal} {Phys. Rev. Lett.}\ }\textbf {\bibinfo {volume} {102}},\
  \bibinfo {pages} {216404} (\bibinfo {year} {2009})}\BibitemShut {NoStop}%
\bibitem [{\citenamefont {Wang}\ \emph
  {et~al.}(2012{\natexlab{a}})\citenamefont {Wang}, \citenamefont {Liu},
  \citenamefont {Xu}, \citenamefont {Yang}, \citenamefont {Miao}, \citenamefont
  {Yao}, \citenamefont {Gao}, \citenamefont {Shen}, \citenamefont {Ma},
  \citenamefont {Chen}, \citenamefont {Xu}, \citenamefont {Liu}, \citenamefont
  {Zhang}, \citenamefont {Qian}, \citenamefont {Jia},\ and\ \citenamefont
  {Xue}}]{Wang2012b}%
  \BibitemOpen
  \bibfield  {author} {\bibinfo {author} {\bibfnamefont {M.-X.}\ \bibnamefont
  {Wang}}, \bibinfo {author} {\bibfnamefont {C.}~\bibnamefont {Liu}}, \bibinfo
  {author} {\bibfnamefont {J.-P.}\ \bibnamefont {Xu}}, \bibinfo {author}
  {\bibfnamefont {F.}~\bibnamefont {Yang}}, \bibinfo {author} {\bibfnamefont
  {L.}~\bibnamefont {Miao}}, \bibinfo {author} {\bibfnamefont {M.-Y.}\
  \bibnamefont {Yao}}, \bibinfo {author} {\bibfnamefont {C.~L.}\ \bibnamefont
  {Gao}}, \bibinfo {author} {\bibfnamefont {C.}~\bibnamefont {Shen}}, \bibinfo
  {author} {\bibfnamefont {X.}~\bibnamefont {Ma}}, \bibinfo {author}
  {\bibfnamefont {X.}~\bibnamefont {Chen}}, \bibinfo {author} {\bibfnamefont
  {Z.-A.}\ \bibnamefont {Xu}}, \bibinfo {author} {\bibfnamefont
  {Y.}~\bibnamefont {Liu}}, \bibinfo {author} {\bibfnamefont {S.-C.}\
  \bibnamefont {Zhang}}, \bibinfo {author} {\bibfnamefont {D.}~\bibnamefont
  {Qian}}, \bibinfo {author} {\bibfnamefont {J.-F.}\ \bibnamefont {Jia}},\ and\
  \bibinfo {author} {\bibfnamefont {Q.-K.}\ \bibnamefont {Xue}},\ }\href
  {https://doi.org/10.1126/science.1216466} {\bibfield  {journal} {\bibinfo
  {journal} {Science}\ }\textbf {\bibinfo {volume} {336}},\ \bibinfo {pages}
  {52} (\bibinfo {year} {2012}{\natexlab{a}})}\BibitemShut {NoStop}%
\bibitem [{\citenamefont {Hosur}\ \emph {et~al.}(2011)\citenamefont {Hosur},
  \citenamefont {Ghaemi}, \citenamefont {Mong},\ and\ \citenamefont
  {Vishwanath}}]{Hosur2011}%
  \BibitemOpen
  \bibfield  {author} {\bibinfo {author} {\bibfnamefont {P.}~\bibnamefont
  {Hosur}}, \bibinfo {author} {\bibfnamefont {P.}~\bibnamefont {Ghaemi}},
  \bibinfo {author} {\bibfnamefont {R.~S.~K.}\ \bibnamefont {Mong}},\ and\
  \bibinfo {author} {\bibfnamefont {A.}~\bibnamefont {Vishwanath}},\ }\href
  {https://doi.org/10.1103/PhysRevLett.107.097001} {\bibfield  {journal}
  {\bibinfo  {journal} {Phys. Rev. Lett.}\ }\textbf {\bibinfo {volume} {107}},\
  \bibinfo {pages} {097001} (\bibinfo {year} {2011})}\BibitemShut {NoStop}%
\bibitem [{\citenamefont {Xu}\ \emph {et~al.}(2014)\citenamefont {Xu},
  \citenamefont {Liu}, \citenamefont {Wang}, \citenamefont {Ge}, \citenamefont
  {Liu}, \citenamefont {Yang}, \citenamefont {Chen}, \citenamefont {Liu},
  \citenamefont {Xu}, \citenamefont {Gao}, \citenamefont {Qian}, \citenamefont
  {Zhang},\ and\ \citenamefont {Jia}}]{Xu2014}%
  \BibitemOpen
  \bibfield  {author} {\bibinfo {author} {\bibfnamefont {J.-P.}\ \bibnamefont
  {Xu}}, \bibinfo {author} {\bibfnamefont {C.}~\bibnamefont {Liu}}, \bibinfo
  {author} {\bibfnamefont {M.-X.}\ \bibnamefont {Wang}}, \bibinfo {author}
  {\bibfnamefont {J.}~\bibnamefont {Ge}}, \bibinfo {author} {\bibfnamefont
  {Z.-L.}\ \bibnamefont {Liu}}, \bibinfo {author} {\bibfnamefont
  {X.}~\bibnamefont {Yang}}, \bibinfo {author} {\bibfnamefont {Y.}~\bibnamefont
  {Chen}}, \bibinfo {author} {\bibfnamefont {Y.}~\bibnamefont {Liu}}, \bibinfo
  {author} {\bibfnamefont {Z.-A.}\ \bibnamefont {Xu}}, \bibinfo {author}
  {\bibfnamefont {C.-L.}\ \bibnamefont {Gao}}, \bibinfo {author} {\bibfnamefont
  {D.}~\bibnamefont {Qian}}, \bibinfo {author} {\bibfnamefont {F.-C.}\
  \bibnamefont {Zhang}},\ and\ \bibinfo {author} {\bibfnamefont {J.-F.}\
  \bibnamefont {Jia}},\ }\href {https://doi.org/10.1103/PhysRevLett.112.217001}
  {\bibfield  {journal} {\bibinfo  {journal} {Phys. Rev. Lett.}\ }\textbf
  {\bibinfo {volume} {112}},\ \bibinfo {pages} {217001} (\bibinfo {year}
  {2014})}\BibitemShut {NoStop}%
\bibitem [{\citenamefont {Hor}\ \emph {et~al.}(2010)\citenamefont {Hor},
  \citenamefont {Williams}, \citenamefont {Checkelsky}, \citenamefont
  {Roushan}, \citenamefont {Seo}, \citenamefont {Xu}, \citenamefont
  {Zandbergen}, \citenamefont {Yazdani}, \citenamefont {Ong},\ and\
  \citenamefont {Cava}}]{Hor2010}%
  \BibitemOpen
  \bibfield  {author} {\bibinfo {author} {\bibfnamefont {Y.~S.}\ \bibnamefont
  {Hor}}, \bibinfo {author} {\bibfnamefont {A.~J.}\ \bibnamefont {Williams}},
  \bibinfo {author} {\bibfnamefont {J.~G.}\ \bibnamefont {Checkelsky}},
  \bibinfo {author} {\bibfnamefont {P.}~\bibnamefont {Roushan}}, \bibinfo
  {author} {\bibfnamefont {J.}~\bibnamefont {Seo}}, \bibinfo {author}
  {\bibfnamefont {Q.}~\bibnamefont {Xu}}, \bibinfo {author} {\bibfnamefont
  {H.~W.}\ \bibnamefont {Zandbergen}}, \bibinfo {author} {\bibfnamefont
  {A.}~\bibnamefont {Yazdani}}, \bibinfo {author} {\bibfnamefont {N.~P.}\
  \bibnamefont {Ong}},\ and\ \bibinfo {author} {\bibfnamefont {R.~J.}\
  \bibnamefont {Cava}},\ }\href
  {https://doi.org/10.1103/PhysRevLett.104.057001} {\bibfield  {journal}
  {\bibinfo  {journal} {Phys. Rev. Lett.}\ }\textbf {\bibinfo {volume} {104}},\
  \bibinfo {pages} {057001} (\bibinfo {year} {2010})}\BibitemShut {NoStop}%
\bibitem [{\citenamefont {Wray}\ \emph {et~al.}(2010)\citenamefont {Wray},
  \citenamefont {Xu}, \citenamefont {Xia}, \citenamefont {Hor}, \citenamefont
  {Qian}, \citenamefont {Fedorov}, \citenamefont {Lin}, \citenamefont {Bansil},
  \citenamefont {Cava},\ and\ \citenamefont {Hasan}}]{Wray2010}%
  \BibitemOpen
  \bibfield  {author} {\bibinfo {author} {\bibfnamefont {L.~A.}\ \bibnamefont
  {Wray}}, \bibinfo {author} {\bibfnamefont {S.-Y.}\ \bibnamefont {Xu}},
  \bibinfo {author} {\bibfnamefont {Y.}~\bibnamefont {Xia}}, \bibinfo {author}
  {\bibfnamefont {Y.~S.}\ \bibnamefont {Hor}}, \bibinfo {author} {\bibfnamefont
  {D.}~\bibnamefont {Qian}}, \bibinfo {author} {\bibfnamefont {A.~V.}\
  \bibnamefont {Fedorov}}, \bibinfo {author} {\bibfnamefont {H.}~\bibnamefont
  {Lin}}, \bibinfo {author} {\bibfnamefont {A.}~\bibnamefont {Bansil}},
  \bibinfo {author} {\bibfnamefont {R.~J.}\ \bibnamefont {Cava}},\ and\
  \bibinfo {author} {\bibfnamefont {M.~Z.}\ \bibnamefont {Hasan}},\ }\href
  {https://doi.org/10.1038/nphys1762} {\bibfield  {journal} {\bibinfo
  {journal} {Nat. Phys.}\ }\textbf {\bibinfo {volume} {6}},\ \bibinfo {pages}
  {855} (\bibinfo {year} {2010})}\BibitemShut {NoStop}%
\bibitem [{\citenamefont {Sasaki}\ \emph {et~al.}(2011)\citenamefont {Sasaki},
  \citenamefont {Kriener}, \citenamefont {Segawa}, \citenamefont {Yada},
  \citenamefont {Tanaka}, \citenamefont {Sato},\ and\ \citenamefont
  {Ando}}]{Sasaki2011}%
  \BibitemOpen
  \bibfield  {author} {\bibinfo {author} {\bibfnamefont {S.}~\bibnamefont
  {Sasaki}}, \bibinfo {author} {\bibfnamefont {M.}~\bibnamefont {Kriener}},
  \bibinfo {author} {\bibfnamefont {K.}~\bibnamefont {Segawa}}, \bibinfo
  {author} {\bibfnamefont {K.}~\bibnamefont {Yada}}, \bibinfo {author}
  {\bibfnamefont {Y.}~\bibnamefont {Tanaka}}, \bibinfo {author} {\bibfnamefont
  {M.}~\bibnamefont {Sato}},\ and\ \bibinfo {author} {\bibfnamefont
  {Y.}~\bibnamefont {Ando}},\ }\href
  {https://doi.org/10.1103/PhysRevLett.107.217001} {\bibfield  {journal}
  {\bibinfo  {journal} {Phys. Rev. Lett.}\ }\textbf {\bibinfo {volume} {107}},\
  \bibinfo {pages} {217001} (\bibinfo {year} {2011})}\BibitemShut {NoStop}%
\bibitem [{\citenamefont {Zhang}\ \emph {et~al.}(2011)\citenamefont {Zhang},
  \citenamefont {Zhang}, \citenamefont {Weng}, \citenamefont {Zhang},
  \citenamefont {Yang}, \citenamefont {Liu}, \citenamefont {Feng},
  \citenamefont {Wang}, \citenamefont {Yu}, \citenamefont {Cao}, \citenamefont
  {Wang}, \citenamefont {Yang}, \citenamefont {Liu}, \citenamefont {Zhao},
  \citenamefont {Zhang}, \citenamefont {Dai}, \citenamefont {Fang},\ and\
  \citenamefont {Jin}}]{Zhang2011}%
  \BibitemOpen
  \bibfield  {author} {\bibinfo {author} {\bibfnamefont {J.~L.}\ \bibnamefont
  {Zhang}}, \bibinfo {author} {\bibfnamefont {S.~J.}\ \bibnamefont {Zhang}},
  \bibinfo {author} {\bibfnamefont {H.~M.}\ \bibnamefont {Weng}}, \bibinfo
  {author} {\bibfnamefont {W.}~\bibnamefont {Zhang}}, \bibinfo {author}
  {\bibfnamefont {L.~X.}\ \bibnamefont {Yang}}, \bibinfo {author}
  {\bibfnamefont {Q.~Q.}\ \bibnamefont {Liu}}, \bibinfo {author} {\bibfnamefont
  {S.~M.}\ \bibnamefont {Feng}}, \bibinfo {author} {\bibfnamefont {X.~C.}\
  \bibnamefont {Wang}}, \bibinfo {author} {\bibfnamefont {R.~C.}\ \bibnamefont
  {Yu}}, \bibinfo {author} {\bibfnamefont {L.~Z.}\ \bibnamefont {Cao}},
  \bibinfo {author} {\bibfnamefont {L.}~\bibnamefont {Wang}}, \bibinfo {author}
  {\bibfnamefont {W.~G.}\ \bibnamefont {Yang}}, \bibinfo {author}
  {\bibfnamefont {H.~Z.}\ \bibnamefont {Liu}}, \bibinfo {author} {\bibfnamefont
  {W.~Y.}\ \bibnamefont {Zhao}}, \bibinfo {author} {\bibfnamefont {S.~C.}\
  \bibnamefont {Zhang}}, \bibinfo {author} {\bibfnamefont {X.}~\bibnamefont
  {Dai}}, \bibinfo {author} {\bibfnamefont {Z.}~\bibnamefont {Fang}},\ and\
  \bibinfo {author} {\bibfnamefont {C.~Q.}\ \bibnamefont {Jin}},\ }\href
  {https://doi.org/10.1073/pnas.1014085108} {\bibfield  {journal} {\bibinfo
  {journal} {Proc. Natl. Acad. Sci. U.S.A.}\ }\textbf {\bibinfo {volume}
  {108}},\ \bibinfo {pages} {24} (\bibinfo {year} {2011})}\BibitemShut
  {NoStop}%
\bibitem [{\citenamefont {Zhu}\ \emph {et~al.}(2013)\citenamefont {Zhu},
  \citenamefont {Zhang}, \citenamefont {Kong}, \citenamefont {Zhang},
  \citenamefont {Yu}, \citenamefont {Zhu}, \citenamefont {Liu}, \citenamefont
  {Li}, \citenamefont {Yu}, \citenamefont {Ahuja}, \citenamefont {Yang},
  \citenamefont {Shen}, \citenamefont {Mao}, \citenamefont {Weng},
  \citenamefont {Dai}, \citenamefont {Fang}, \citenamefont {Zhao},\ and\
  \citenamefont {Jin}}]{Zhu2013}%
  \BibitemOpen
  \bibfield  {author} {\bibinfo {author} {\bibfnamefont {J.}~\bibnamefont
  {Zhu}}, \bibinfo {author} {\bibfnamefont {J.~L.}\ \bibnamefont {Zhang}},
  \bibinfo {author} {\bibfnamefont {P.~P.}\ \bibnamefont {Kong}}, \bibinfo
  {author} {\bibfnamefont {S.~J.}\ \bibnamefont {Zhang}}, \bibinfo {author}
  {\bibfnamefont {X.~H.}\ \bibnamefont {Yu}}, \bibinfo {author} {\bibfnamefont
  {J.~L.}\ \bibnamefont {Zhu}}, \bibinfo {author} {\bibfnamefont {Q.~Q.}\
  \bibnamefont {Liu}}, \bibinfo {author} {\bibfnamefont {X.}~\bibnamefont
  {Li}}, \bibinfo {author} {\bibfnamefont {R.~C.}\ \bibnamefont {Yu}}, \bibinfo
  {author} {\bibfnamefont {R.}~\bibnamefont {Ahuja}}, \bibinfo {author}
  {\bibfnamefont {W.~G.}\ \bibnamefont {Yang}}, \bibinfo {author}
  {\bibfnamefont {G.~Y.}\ \bibnamefont {Shen}}, \bibinfo {author}
  {\bibfnamefont {H.~K.}\ \bibnamefont {Mao}}, \bibinfo {author} {\bibfnamefont
  {H.~M.}\ \bibnamefont {Weng}}, \bibinfo {author} {\bibfnamefont
  {X.}~\bibnamefont {Dai}}, \bibinfo {author} {\bibfnamefont {Z.}~\bibnamefont
  {Fang}}, \bibinfo {author} {\bibfnamefont {Y.~S.}\ \bibnamefont {Zhao}},\
  and\ \bibinfo {author} {\bibfnamefont {C.~Q.}\ \bibnamefont {Jin}},\ }\href
  {https://doi.org/10.1038/srep02016} {\bibfield  {journal} {\bibinfo
  {journal} {Sci. Rep.}\ }\textbf {\bibinfo {volume} {3}},\ \bibinfo {pages}
  {2016} (\bibinfo {year} {2013})}\BibitemShut {NoStop}%
\bibitem [{\citenamefont {Kirshenbaum}\ \emph {et~al.}(2013)\citenamefont
  {Kirshenbaum}, \citenamefont {Syers}, \citenamefont {Hope}, \citenamefont
  {Butch}, \citenamefont {Jeffries}, \citenamefont {Weir}, \citenamefont
  {Hamlin}, \citenamefont {Maple}, \citenamefont {Vohra},\ and\ \citenamefont
  {Paglione}}]{Kirshenbaum2013}%
  \BibitemOpen
  \bibfield  {author} {\bibinfo {author} {\bibfnamefont {K.}~\bibnamefont
  {Kirshenbaum}}, \bibinfo {author} {\bibfnamefont {P.~S.}\ \bibnamefont
  {Syers}}, \bibinfo {author} {\bibfnamefont {A.~P.}\ \bibnamefont {Hope}},
  \bibinfo {author} {\bibfnamefont {N.~P.}\ \bibnamefont {Butch}}, \bibinfo
  {author} {\bibfnamefont {J.~R.}\ \bibnamefont {Jeffries}}, \bibinfo {author}
  {\bibfnamefont {S.~T.}\ \bibnamefont {Weir}}, \bibinfo {author}
  {\bibfnamefont {J.~J.}\ \bibnamefont {Hamlin}}, \bibinfo {author}
  {\bibfnamefont {M.~B.}\ \bibnamefont {Maple}}, \bibinfo {author}
  {\bibfnamefont {Y.~K.}\ \bibnamefont {Vohra}},\ and\ \bibinfo {author}
  {\bibfnamefont {J.}~\bibnamefont {Paglione}},\ }\href
  {https://doi.org/10.1103/PhysRevLett.111.087001} {\bibfield  {journal}
  {\bibinfo  {journal} {Phys. Rev. Lett.}\ }\textbf {\bibinfo {volume} {111}},\
  \bibinfo {pages} {087001} (\bibinfo {year} {2013})}\BibitemShut {NoStop}%
\bibitem [{\citenamefont {Shruti}\ \emph {et~al.}(2015)\citenamefont {Shruti},
  \citenamefont {Maurya}, \citenamefont {Neha}, \citenamefont {Srivastava},\
  and\ \citenamefont {Patnaik}}]{Shruti2015}%
  \BibitemOpen
  \bibfield  {author} {\bibinfo {author} {\bibnamefont {Shruti}}, \bibinfo
  {author} {\bibfnamefont {V.~K.}\ \bibnamefont {Maurya}}, \bibinfo {author}
  {\bibfnamefont {P.}~\bibnamefont {Neha}}, \bibinfo {author} {\bibfnamefont
  {P.}~\bibnamefont {Srivastava}},\ and\ \bibinfo {author} {\bibfnamefont
  {S.}~\bibnamefont {Patnaik}},\ }\href
  {https://doi.org/10.1103/PhysRevB.92.020506} {\bibfield  {journal} {\bibinfo
  {journal} {Phys. Rev. B}\ }\textbf {\bibinfo {volume} {92}},\ \bibinfo
  {pages} {020506} (\bibinfo {year} {2015})}\BibitemShut {NoStop}%
\bibitem [{\citenamefont {He}\ \emph {et~al.}(2016)\citenamefont {He},
  \citenamefont {Jia}, \citenamefont {Zhang}, \citenamefont {Hong},
  \citenamefont {Jin},\ and\ \citenamefont {Li}}]{He2016}%
  \BibitemOpen
  \bibfield  {author} {\bibinfo {author} {\bibfnamefont {L.}~\bibnamefont
  {He}}, \bibinfo {author} {\bibfnamefont {Y.}~\bibnamefont {Jia}}, \bibinfo
  {author} {\bibfnamefont {S.}~\bibnamefont {Zhang}}, \bibinfo {author}
  {\bibfnamefont {X.}~\bibnamefont {Hong}}, \bibinfo {author} {\bibfnamefont
  {C.}~\bibnamefont {Jin}},\ and\ \bibinfo {author} {\bibfnamefont
  {S.}~\bibnamefont {Li}},\ }\href
  {https://doi.org/10.1038/npjquantmats.2016.14} {\bibfield  {journal}
  {\bibinfo  {journal} {npj Quantum Mater.}\ }\textbf {\bibinfo {volume} {1}},\
  \bibinfo {pages} {16014} (\bibinfo {year} {2016})}\BibitemShut {NoStop}%
\bibitem [{\citenamefont {Zhou}\ \emph {et~al.}(2016)\citenamefont {Zhou},
  \citenamefont {Wu}, \citenamefont {Ning}, \citenamefont {Li}, \citenamefont
  {Du}, \citenamefont {Chen}, \citenamefont {Zhang}, \citenamefont {Chi},
  \citenamefont {Wang}, \citenamefont {Zhu}, \citenamefont {Lu}, \citenamefont
  {Ji}, \citenamefont {Wan}, \citenamefont {Yang}, \citenamefont {Sun},
  \citenamefont {Yang}, \citenamefont {Tian}, \citenamefont {Zhang},\ and\
  \citenamefont {Mao}}]{Zhou2016}%
  \BibitemOpen
  \bibfield  {author} {\bibinfo {author} {\bibfnamefont {Y.}~\bibnamefont
  {Zhou}}, \bibinfo {author} {\bibfnamefont {J.}~\bibnamefont {Wu}}, \bibinfo
  {author} {\bibfnamefont {W.}~\bibnamefont {Ning}}, \bibinfo {author}
  {\bibfnamefont {N.}~\bibnamefont {Li}}, \bibinfo {author} {\bibfnamefont
  {Y.}~\bibnamefont {Du}}, \bibinfo {author} {\bibfnamefont {X.}~\bibnamefont
  {Chen}}, \bibinfo {author} {\bibfnamefont {R.}~\bibnamefont {Zhang}},
  \bibinfo {author} {\bibfnamefont {Z.}~\bibnamefont {Chi}}, \bibinfo {author}
  {\bibfnamefont {X.}~\bibnamefont {Wang}}, \bibinfo {author} {\bibfnamefont
  {X.}~\bibnamefont {Zhu}}, \bibinfo {author} {\bibfnamefont {P.}~\bibnamefont
  {Lu}}, \bibinfo {author} {\bibfnamefont {C.}~\bibnamefont {Ji}}, \bibinfo
  {author} {\bibfnamefont {X.}~\bibnamefont {Wan}}, \bibinfo {author}
  {\bibfnamefont {Z.}~\bibnamefont {Yang}}, \bibinfo {author} {\bibfnamefont
  {J.}~\bibnamefont {Sun}}, \bibinfo {author} {\bibfnamefont {W.}~\bibnamefont
  {Yang}}, \bibinfo {author} {\bibfnamefont {M.}~\bibnamefont {Tian}}, \bibinfo
  {author} {\bibfnamefont {Y.}~\bibnamefont {Zhang}},\ and\ \bibinfo {author}
  {\bibfnamefont {H.-k.}\ \bibnamefont {Mao}},\ }\href
  {https://doi.org/10.1073/pnas.1601262113} {\bibfield  {journal} {\bibinfo
  {journal} {Proc. Natl. Acad. Sci. U.S.A.}\ }\textbf {\bibinfo {volume}
  {113}},\ \bibinfo {pages} {2904} (\bibinfo {year} {2016})}\BibitemShut
  {NoStop}%
\bibitem [{\citenamefont {Chen}\ \emph {et~al.}(2009)\citenamefont {Chen},
  \citenamefont {Analytis}, \citenamefont {Chu}, \citenamefont {Liu},
  \citenamefont {Mo}, \citenamefont {Qi}, \citenamefont {Zhang}, \citenamefont
  {Lu}, \citenamefont {Dai}, \citenamefont {Fang}, \citenamefont {Zhang},
  \citenamefont {Fisher}, \citenamefont {Hussain},\ and\ \citenamefont
  {Shen}}]{Chen2009}%
  \BibitemOpen
  \bibfield  {author} {\bibinfo {author} {\bibfnamefont {Y.~L.}\ \bibnamefont
  {Chen}}, \bibinfo {author} {\bibfnamefont {J.~G.}\ \bibnamefont {Analytis}},
  \bibinfo {author} {\bibfnamefont {J.-H.}\ \bibnamefont {Chu}}, \bibinfo
  {author} {\bibfnamefont {Z.~K.}\ \bibnamefont {Liu}}, \bibinfo {author}
  {\bibfnamefont {S.-K.}\ \bibnamefont {Mo}}, \bibinfo {author} {\bibfnamefont
  {X.~L.}\ \bibnamefont {Qi}}, \bibinfo {author} {\bibfnamefont {H.~J.}\
  \bibnamefont {Zhang}}, \bibinfo {author} {\bibfnamefont {D.~H.}\ \bibnamefont
  {Lu}}, \bibinfo {author} {\bibfnamefont {X.}~\bibnamefont {Dai}}, \bibinfo
  {author} {\bibfnamefont {Z.}~\bibnamefont {Fang}}, \bibinfo {author}
  {\bibfnamefont {S.~C.}\ \bibnamefont {Zhang}}, \bibinfo {author}
  {\bibfnamefont {I.~R.}\ \bibnamefont {Fisher}}, \bibinfo {author}
  {\bibfnamefont {Z.}~\bibnamefont {Hussain}},\ and\ \bibinfo {author}
  {\bibfnamefont {Z.-X.}\ \bibnamefont {Shen}},\ }\href
  {https://doi.org/10.1126/science.1173034} {\bibfield  {journal} {\bibinfo
  {journal} {Science}\ }\textbf {\bibinfo {volume} {325}},\ \bibinfo {pages}
  {178} (\bibinfo {year} {2009})}\BibitemShut {NoStop}%
\bibitem [{\citenamefont {Zhang}\ \emph {et~al.}(2009)\citenamefont {Zhang},
  \citenamefont {Liu}, \citenamefont {Qi}, \citenamefont {Dai}, \citenamefont
  {Fang},\ and\ \citenamefont {Zhang}}]{Zhang2009}%
  \BibitemOpen
  \bibfield  {author} {\bibinfo {author} {\bibfnamefont {H.}~\bibnamefont
  {Zhang}}, \bibinfo {author} {\bibfnamefont {C.-X.}\ \bibnamefont {Liu}},
  \bibinfo {author} {\bibfnamefont {X.-L.}\ \bibnamefont {Qi}}, \bibinfo
  {author} {\bibfnamefont {X.}~\bibnamefont {Dai}}, \bibinfo {author}
  {\bibfnamefont {Z.}~\bibnamefont {Fang}},\ and\ \bibinfo {author}
  {\bibfnamefont {S.-C.}\ \bibnamefont {Zhang}},\ }\href
  {https://doi.org/10.1038/nphys1270} {\bibfield  {journal} {\bibinfo
  {journal} {Nat. Phys.}\ }\textbf {\bibinfo {volume} {5}},\ \bibinfo {pages}
  {438} (\bibinfo {year} {2009})}\BibitemShut {NoStop}%
\bibitem [{\citenamefont {Hasan}\ and\ \citenamefont {Kane}(2010)}]{Hasan2010}%
  \BibitemOpen
  \bibfield  {author} {\bibinfo {author} {\bibfnamefont {M.~Z.}\ \bibnamefont
  {Hasan}}\ and\ \bibinfo {author} {\bibfnamefont {C.~L.}\ \bibnamefont
  {Kane}},\ }\href {https://doi.org/10.1103/RevModPhys.82.3045} {\bibfield
  {journal} {\bibinfo  {journal} {Rev. Mod. Phys.}\ }\textbf {\bibinfo {volume}
  {82}},\ \bibinfo {pages} {3045} (\bibinfo {year} {2010})}\BibitemShut
  {NoStop}%
\bibitem [{\citenamefont {Hasan}\ and\ \citenamefont
  {Moore}(2011)}]{Hasan2011}%
  \BibitemOpen
  \bibfield  {author} {\bibinfo {author} {\bibfnamefont {M.~Z.}\ \bibnamefont
  {Hasan}}\ and\ \bibinfo {author} {\bibfnamefont {J.~E.}\ \bibnamefont
  {Moore}},\ }\href {https://doi.org/10.1146/annurev-conmatphys-062910-140432}
  {\bibfield  {journal} {\bibinfo  {journal} {Annu. Rev. Condens. Matter
  Phys.}\ }\textbf {\bibinfo {volume} {2}},\ \bibinfo {pages} {55} (\bibinfo
  {year} {2011})}\BibitemShut {NoStop}%
\bibitem [{\citenamefont {Wang}\ \emph
  {et~al.}(2012{\natexlab{b}})\citenamefont {Wang}, \citenamefont {Sun},
  \citenamefont {Chen}, \citenamefont {Franchini}, \citenamefont {Xu},
  \citenamefont {Weng}, \citenamefont {Dai},\ and\ \citenamefont
  {Fang}}]{Wang2012}%
  \BibitemOpen
  \bibfield  {author} {\bibinfo {author} {\bibfnamefont {Z.}~\bibnamefont
  {Wang}}, \bibinfo {author} {\bibfnamefont {Y.}~\bibnamefont {Sun}}, \bibinfo
  {author} {\bibfnamefont {X.-Q.}\ \bibnamefont {Chen}}, \bibinfo {author}
  {\bibfnamefont {C.}~\bibnamefont {Franchini}}, \bibinfo {author}
  {\bibfnamefont {G.}~\bibnamefont {Xu}}, \bibinfo {author} {\bibfnamefont
  {H.}~\bibnamefont {Weng}}, \bibinfo {author} {\bibfnamefont {X.}~\bibnamefont
  {Dai}},\ and\ \bibinfo {author} {\bibfnamefont {Z.}~\bibnamefont {Fang}},\
  }\href {https://doi.org/10.1103/PhysRevB.85.195320} {\bibfield  {journal}
  {\bibinfo  {journal} {Phys. Rev. B}\ }\textbf {\bibinfo {volume} {85}},\
  \bibinfo {pages} {195320} (\bibinfo {year} {2012}{\natexlab{b}})}\BibitemShut
  {NoStop}%
\bibitem [{\citenamefont {Weng}\ \emph {et~al.}(2015)\citenamefont {Weng},
  \citenamefont {Fang}, \citenamefont {Fang}, \citenamefont {Bernevig},\ and\
  \citenamefont {Dai}}]{Weng2015}%
  \BibitemOpen
  \bibfield  {author} {\bibinfo {author} {\bibfnamefont {H.}~\bibnamefont
  {Weng}}, \bibinfo {author} {\bibfnamefont {C.}~\bibnamefont {Fang}}, \bibinfo
  {author} {\bibfnamefont {Z.}~\bibnamefont {Fang}}, \bibinfo {author}
  {\bibfnamefont {B.~A.}\ \bibnamefont {Bernevig}},\ and\ \bibinfo {author}
  {\bibfnamefont {X.}~\bibnamefont {Dai}},\ }\href
  {https://doi.org/10.1103/PhysRevX.5.011029} {\bibfield  {journal} {\bibinfo
  {journal} {Phys. Rev. X}\ }\textbf {\bibinfo {volume} {5}},\ \bibinfo {pages}
  {011029} (\bibinfo {year} {2015})}\BibitemShut {NoStop}%
\bibitem [{\citenamefont {Armitage}\ \emph {et~al.}(2018)\citenamefont
  {Armitage}, \citenamefont {Mele},\ and\ \citenamefont
  {Vishwanath}}]{Armitage2018}%
  \BibitemOpen
  \bibfield  {author} {\bibinfo {author} {\bibfnamefont {N.~P.}\ \bibnamefont
  {Armitage}}, \bibinfo {author} {\bibfnamefont {E.~J.}\ \bibnamefont {Mele}},\
  and\ \bibinfo {author} {\bibfnamefont {A.}~\bibnamefont {Vishwanath}},\
  }\href {https://doi.org/10.1103/RevModPhys.90.015001} {\bibfield  {journal}
  {\bibinfo  {journal} {Rev. Mod. Phys.}\ }\textbf {\bibinfo {volume} {90}},\
  \bibinfo {pages} {015001} (\bibinfo {year} {2018})}\BibitemShut {NoStop}%
\bibitem [{\citenamefont {Soluyanov}\ \emph {et~al.}(2015)\citenamefont
  {Soluyanov}, \citenamefont {Gresch}, \citenamefont {Wang}, \citenamefont
  {Wu}, \citenamefont {Troyer}, \citenamefont {Dai},\ and\ \citenamefont
  {Bernevig}}]{Soluyanov2015}%
  \BibitemOpen
  \bibfield  {author} {\bibinfo {author} {\bibfnamefont {A.~A.}\ \bibnamefont
  {Soluyanov}}, \bibinfo {author} {\bibfnamefont {D.}~\bibnamefont {Gresch}},
  \bibinfo {author} {\bibfnamefont {Z.}~\bibnamefont {Wang}}, \bibinfo {author}
  {\bibfnamefont {Q.}~\bibnamefont {Wu}}, \bibinfo {author} {\bibfnamefont
  {M.}~\bibnamefont {Troyer}}, \bibinfo {author} {\bibfnamefont
  {X.}~\bibnamefont {Dai}},\ and\ \bibinfo {author} {\bibfnamefont {B.~A.}\
  \bibnamefont {Bernevig}},\ }\href {https://doi.org/10.1038/nature15768}
  {\bibfield  {journal} {\bibinfo  {journal} {Nature}\ }\textbf {\bibinfo
  {volume} {527}},\ \bibinfo {pages} {495} (\bibinfo {year}
  {2015})}\BibitemShut {NoStop}%
\bibitem [{\citenamefont {Chiu}\ \emph {et~al.}(2016)\citenamefont {Chiu},
  \citenamefont {Teo}, \citenamefont {Schnyder},\ and\ \citenamefont
  {Ryu}}]{Chiu2016}%
  \BibitemOpen
  \bibfield  {author} {\bibinfo {author} {\bibfnamefont {C.-K.}\ \bibnamefont
  {Chiu}}, \bibinfo {author} {\bibfnamefont {J.~C.~Y.}\ \bibnamefont {Teo}},
  \bibinfo {author} {\bibfnamefont {A.~P.}\ \bibnamefont {Schnyder}},\ and\
  \bibinfo {author} {\bibfnamefont {S.}~\bibnamefont {Ryu}},\ }\href
  {https://doi.org/10.1103/RevModPhys.88.035005} {\bibfield  {journal}
  {\bibinfo  {journal} {Rev. Mod. Phys.}\ }\textbf {\bibinfo {volume} {88}},\
  \bibinfo {pages} {035005} (\bibinfo {year} {2016})}\BibitemShut {NoStop}%
\bibitem [{\citenamefont {Young}\ \emph {et~al.}(2012)\citenamefont {Young},
  \citenamefont {Zaheer}, \citenamefont {Teo}, \citenamefont {Kane},
  \citenamefont {Mele},\ and\ \citenamefont {Rappe}}]{Young2012}%
  \BibitemOpen
  \bibfield  {author} {\bibinfo {author} {\bibfnamefont {S.~M.}\ \bibnamefont
  {Young}}, \bibinfo {author} {\bibfnamefont {S.}~\bibnamefont {Zaheer}},
  \bibinfo {author} {\bibfnamefont {J.~C.~Y.}\ \bibnamefont {Teo}}, \bibinfo
  {author} {\bibfnamefont {C.~L.}\ \bibnamefont {Kane}}, \bibinfo {author}
  {\bibfnamefont {E.~J.}\ \bibnamefont {Mele}},\ and\ \bibinfo {author}
  {\bibfnamefont {A.~M.}\ \bibnamefont {Rappe}},\ }\href
  {https://doi.org/10.1103/PhysRevLett.108.140405} {\bibfield  {journal}
  {\bibinfo  {journal} {Phys. Rev. Lett.}\ }\textbf {\bibinfo {volume} {108}},\
  \bibinfo {pages} {140405} (\bibinfo {year} {2012})}\BibitemShut {NoStop}%
\bibitem [{\citenamefont {Lv}\ \emph {et~al.}(2015)\citenamefont {Lv},
  \citenamefont {Weng}, \citenamefont {Fu}, \citenamefont {Wang}, \citenamefont
  {Miao}, \citenamefont {Ma}, \citenamefont {Richard}, \citenamefont {Huang},
  \citenamefont {Zhao}, \citenamefont {Chen}, \citenamefont {Fang},
  \citenamefont {Dai}, \citenamefont {Qian},\ and\ \citenamefont
  {Ding}}]{Lv2015}%
  \BibitemOpen
  \bibfield  {author} {\bibinfo {author} {\bibfnamefont {B.~Q.}\ \bibnamefont
  {Lv}}, \bibinfo {author} {\bibfnamefont {H.~M.}\ \bibnamefont {Weng}},
  \bibinfo {author} {\bibfnamefont {B.~B.}\ \bibnamefont {Fu}}, \bibinfo
  {author} {\bibfnamefont {X.~P.}\ \bibnamefont {Wang}}, \bibinfo {author}
  {\bibfnamefont {H.}~\bibnamefont {Miao}}, \bibinfo {author} {\bibfnamefont
  {J.}~\bibnamefont {Ma}}, \bibinfo {author} {\bibfnamefont {P.}~\bibnamefont
  {Richard}}, \bibinfo {author} {\bibfnamefont {X.~C.}\ \bibnamefont {Huang}},
  \bibinfo {author} {\bibfnamefont {L.~X.}\ \bibnamefont {Zhao}}, \bibinfo
  {author} {\bibfnamefont {G.~F.}\ \bibnamefont {Chen}}, \bibinfo {author}
  {\bibfnamefont {Z.}~\bibnamefont {Fang}}, \bibinfo {author} {\bibfnamefont
  {X.}~\bibnamefont {Dai}}, \bibinfo {author} {\bibfnamefont {T.}~\bibnamefont
  {Qian}},\ and\ \bibinfo {author} {\bibfnamefont {H.}~\bibnamefont {Ding}},\
  }\href {https://doi.org/10.1103/PhysRevX.5.031013} {\bibfield  {journal}
  {\bibinfo  {journal} {Phys. Rev. X}\ }\textbf {\bibinfo {volume} {5}},\
  \bibinfo {pages} {031013} (\bibinfo {year} {2015})}\BibitemShut {NoStop}%
\bibitem [{\citenamefont {Schoop}\ \emph {et~al.}(2016)\citenamefont {Schoop},
  \citenamefont {Ali}, \citenamefont {Stra\ss{}er}, \citenamefont {Topp},
  \citenamefont {Varykhalov}, \citenamefont {Marchenko}, \citenamefont
  {Duppel}, \citenamefont {Parkin}, \citenamefont {Lotsch},\ and\ \citenamefont
  {Ast}}]{Schoop2016}%
  \BibitemOpen
  \bibfield  {author} {\bibinfo {author} {\bibfnamefont {L.~M.}\ \bibnamefont
  {Schoop}}, \bibinfo {author} {\bibfnamefont {M.~N.}\ \bibnamefont {Ali}},
  \bibinfo {author} {\bibfnamefont {C.}~\bibnamefont {Stra\ss{}er}}, \bibinfo
  {author} {\bibfnamefont {A.}~\bibnamefont {Topp}}, \bibinfo {author}
  {\bibfnamefont {A.}~\bibnamefont {Varykhalov}}, \bibinfo {author}
  {\bibfnamefont {D.}~\bibnamefont {Marchenko}}, \bibinfo {author}
  {\bibfnamefont {V.}~\bibnamefont {Duppel}}, \bibinfo {author} {\bibfnamefont
  {S.~S.~P.}\ \bibnamefont {Parkin}}, \bibinfo {author} {\bibfnamefont {B.~V.}\
  \bibnamefont {Lotsch}},\ and\ \bibinfo {author} {\bibfnamefont {C.~R.}\
  \bibnamefont {Ast}},\ }\href {https://doi.org/10.1038/ncomms11696} {\bibfield
   {journal} {\bibinfo  {journal} {Nat. Commun.}\ }\textbf {\bibinfo {volume}
  {7}},\ \bibinfo {pages} {11696} (\bibinfo {year} {2016})}\BibitemShut
  {NoStop}%
\bibitem [{\citenamefont {Zhang}\ \emph {et~al.}(2018)\citenamefont {Zhang},
  \citenamefont {Yu}, \citenamefont {Zhu}, \citenamefont {Wu}, \citenamefont
  {Wang}, \citenamefont {Sheng},\ and\ \citenamefont {Yang}}]{Zhang2018}%
  \BibitemOpen
  \bibfield  {author} {\bibinfo {author} {\bibfnamefont {X.}~\bibnamefont
  {Zhang}}, \bibinfo {author} {\bibfnamefont {Z.-M.}\ \bibnamefont {Yu}},
  \bibinfo {author} {\bibfnamefont {Z.}~\bibnamefont {Zhu}}, \bibinfo {author}
  {\bibfnamefont {W.}~\bibnamefont {Wu}}, \bibinfo {author} {\bibfnamefont
  {S.-S.}\ \bibnamefont {Wang}}, \bibinfo {author} {\bibfnamefont {X.-L.}\
  \bibnamefont {Sheng}},\ and\ \bibinfo {author} {\bibfnamefont {S.~A.}\
  \bibnamefont {Yang}},\ }\href {https://doi.org/10.1103/PhysRevB.97.235150}
  {\bibfield  {journal} {\bibinfo  {journal} {Phys. Rev. B}\ }\textbf {\bibinfo
  {volume} {97}},\ \bibinfo {pages} {235150} (\bibinfo {year}
  {2018})}\BibitemShut {NoStop}%
\bibitem [{\citenamefont {Weng}\ \emph {et~al.}(2016)\citenamefont {Weng},
  \citenamefont {Fang}, \citenamefont {Fang},\ and\ \citenamefont
  {Dai}}]{Weng2016}%
  \BibitemOpen
  \bibfield  {author} {\bibinfo {author} {\bibfnamefont {H.}~\bibnamefont
  {Weng}}, \bibinfo {author} {\bibfnamefont {C.}~\bibnamefont {Fang}}, \bibinfo
  {author} {\bibfnamefont {Z.}~\bibnamefont {Fang}},\ and\ \bibinfo {author}
  {\bibfnamefont {X.}~\bibnamefont {Dai}},\ }\href
  {https://doi.org/10.1103/PhysRevB.93.241202} {\bibfield  {journal} {\bibinfo
  {journal} {Phys. Rev. B}\ }\textbf {\bibinfo {volume} {93}},\ \bibinfo
  {pages} {241202} (\bibinfo {year} {2016})}\BibitemShut {NoStop}%
\bibitem [{\citenamefont {Burkov}\ \emph {et~al.}(2011)\citenamefont {Burkov},
  \citenamefont {Hook},\ and\ \citenamefont {Balents}}]{Burkov2011}%
  \BibitemOpen
  \bibfield  {author} {\bibinfo {author} {\bibfnamefont {A.~A.}\ \bibnamefont
  {Burkov}}, \bibinfo {author} {\bibfnamefont {M.~D.}\ \bibnamefont {Hook}},\
  and\ \bibinfo {author} {\bibfnamefont {L.}~\bibnamefont {Balents}},\ }\href
  {https://doi.org/10.1103/PhysRevB.84.235126} {\bibfield  {journal} {\bibinfo
  {journal} {Phys. Rev. B}\ }\textbf {\bibinfo {volume} {84}},\ \bibinfo
  {pages} {235126} (\bibinfo {year} {2011})}\BibitemShut {NoStop}%
\bibitem [{\citenamefont {Lau}\ \emph {et~al.}(2021)\citenamefont {Lau},
  \citenamefont {Hyart}, \citenamefont {Autieri}, \citenamefont {Chen},\ and\
  \citenamefont {Pikulin}}]{Lau2021}%
  \BibitemOpen
  \bibfield  {author} {\bibinfo {author} {\bibfnamefont {A.}~\bibnamefont
  {Lau}}, \bibinfo {author} {\bibfnamefont {T.}~\bibnamefont {Hyart}}, \bibinfo
  {author} {\bibfnamefont {C.}~\bibnamefont {Autieri}}, \bibinfo {author}
  {\bibfnamefont {A.}~\bibnamefont {Chen}},\ and\ \bibinfo {author}
  {\bibfnamefont {D.~I.}\ \bibnamefont {Pikulin}},\ }\href
  {https://doi.org/10.1103/PhysRevX.11.031017} {\bibfield  {journal} {\bibinfo
  {journal} {Phys. Rev. X}\ }\textbf {\bibinfo {volume} {11}},\ \bibinfo
  {pages} {031017} (\bibinfo {year} {2021})}\BibitemShut {NoStop}%
\bibitem [{\citenamefont {Ali}\ \emph {et~al.}(2014)\citenamefont {Ali},
  \citenamefont {Gibson}, \citenamefont {Klimczuk},\ and\ \citenamefont
  {Cava}}]{Ali2014}%
  \BibitemOpen
  \bibfield  {author} {\bibinfo {author} {\bibfnamefont {M.~N.}\ \bibnamefont
  {Ali}}, \bibinfo {author} {\bibfnamefont {Q.~D.}\ \bibnamefont {Gibson}},
  \bibinfo {author} {\bibfnamefont {T.}~\bibnamefont {Klimczuk}},\ and\
  \bibinfo {author} {\bibfnamefont {R.~J.}\ \bibnamefont {Cava}},\ }\href
  {https://doi.org/10.1103/PhysRevB.89.020505} {\bibfield  {journal} {\bibinfo
  {journal} {Phys. Rev. B}\ }\textbf {\bibinfo {volume} {89}},\ \bibinfo
  {pages} {020505} (\bibinfo {year} {2014})}\BibitemShut {NoStop}%
\bibitem [{\citenamefont {Bian}\ \emph {et~al.}(2016)\citenamefont {Bian},
  \citenamefont {Chang}, \citenamefont {Sankar}, \citenamefont {Xu},
  \citenamefont {Zheng}, \citenamefont {Neupert}, \citenamefont {Chiu},
  \citenamefont {Huang}, \citenamefont {Chang}, \citenamefont {Belopolski},
  \citenamefont {Sanchez}, \citenamefont {Neupane}, \citenamefont {Alidoust},
  \citenamefont {Liu}, \citenamefont {Wang}, \citenamefont {Lee}, \citenamefont
  {Jeng}, \citenamefont {Zhang}, \citenamefont {Yuan}, \citenamefont {Jia},
  \citenamefont {Bansil}, \citenamefont {Chou}, \citenamefont {Lin},\ and\
  \citenamefont {Hasan}}]{Bian2016}%
  \BibitemOpen
  \bibfield  {author} {\bibinfo {author} {\bibfnamefont {G.}~\bibnamefont
  {Bian}}, \bibinfo {author} {\bibfnamefont {T.-R.}\ \bibnamefont {Chang}},
  \bibinfo {author} {\bibfnamefont {R.}~\bibnamefont {Sankar}}, \bibinfo
  {author} {\bibfnamefont {S.-Y.}\ \bibnamefont {Xu}}, \bibinfo {author}
  {\bibfnamefont {H.}~\bibnamefont {Zheng}}, \bibinfo {author} {\bibfnamefont
  {T.}~\bibnamefont {Neupert}}, \bibinfo {author} {\bibfnamefont {C.-K.}\
  \bibnamefont {Chiu}}, \bibinfo {author} {\bibfnamefont {S.-M.}\ \bibnamefont
  {Huang}}, \bibinfo {author} {\bibfnamefont {G.}~\bibnamefont {Chang}},
  \bibinfo {author} {\bibfnamefont {I.}~\bibnamefont {Belopolski}}, \bibinfo
  {author} {\bibfnamefont {D.~S.}\ \bibnamefont {Sanchez}}, \bibinfo {author}
  {\bibfnamefont {M.}~\bibnamefont {Neupane}}, \bibinfo {author} {\bibfnamefont
  {N.}~\bibnamefont {Alidoust}}, \bibinfo {author} {\bibfnamefont
  {C.}~\bibnamefont {Liu}}, \bibinfo {author} {\bibfnamefont {B.}~\bibnamefont
  {Wang}}, \bibinfo {author} {\bibfnamefont {C.-C.}\ \bibnamefont {Lee}},
  \bibinfo {author} {\bibfnamefont {H.-T.}\ \bibnamefont {Jeng}}, \bibinfo
  {author} {\bibfnamefont {C.}~\bibnamefont {Zhang}}, \bibinfo {author}
  {\bibfnamefont {Z.}~\bibnamefont {Yuan}}, \bibinfo {author} {\bibfnamefont
  {S.}~\bibnamefont {Jia}}, \bibinfo {author} {\bibfnamefont {A.}~\bibnamefont
  {Bansil}}, \bibinfo {author} {\bibfnamefont {F.}~\bibnamefont {Chou}},
  \bibinfo {author} {\bibfnamefont {H.}~\bibnamefont {Lin}},\ and\ \bibinfo
  {author} {\bibfnamefont {M.~Z.}\ \bibnamefont {Hasan}},\ }\href
  {https://doi.org/10.1038/ncomms10556} {\bibfield  {journal} {\bibinfo
  {journal} {Nat. Commun.}\ }\textbf {\bibinfo {volume} {7}},\ \bibinfo {pages}
  {10556} (\bibinfo {year} {2016})}\BibitemShut {NoStop}%
\bibitem [{\citenamefont {Zhang}\ \emph {et~al.}(2016)\citenamefont {Zhang},
  \citenamefont {Yuan}, \citenamefont {Bian}, \citenamefont {Xu}, \citenamefont
  {Zhang}, \citenamefont {Hasan},\ and\ \citenamefont {Jia}}]{Zhang2016}%
  \BibitemOpen
  \bibfield  {author} {\bibinfo {author} {\bibfnamefont {C.-L.}\ \bibnamefont
  {Zhang}}, \bibinfo {author} {\bibfnamefont {Z.}~\bibnamefont {Yuan}},
  \bibinfo {author} {\bibfnamefont {G.}~\bibnamefont {Bian}}, \bibinfo {author}
  {\bibfnamefont {S.-Y.}\ \bibnamefont {Xu}}, \bibinfo {author} {\bibfnamefont
  {X.}~\bibnamefont {Zhang}}, \bibinfo {author} {\bibfnamefont {M.~Z.}\
  \bibnamefont {Hasan}},\ and\ \bibinfo {author} {\bibfnamefont
  {S.}~\bibnamefont {Jia}},\ }\href
  {https://doi.org/10.1103/PhysRevB.93.054520} {\bibfield  {journal} {\bibinfo
  {journal} {Phys. Rev. B}\ }\textbf {\bibinfo {volume} {93}},\ \bibinfo
  {pages} {054520} (\bibinfo {year} {2016})}\BibitemShut {NoStop}%
\bibitem [{\citenamefont {Guan}\ \emph {et~al.}(2016)\citenamefont {Guan},
  \citenamefont {Chen}, \citenamefont {Chu}, \citenamefont {Sankar},
  \citenamefont {Chou}, \citenamefont {Jeng}, \citenamefont {Chang},\ and\
  \citenamefont {Chuang}}]{Guan2016}%
  \BibitemOpen
  \bibfield  {author} {\bibinfo {author} {\bibfnamefont {S.-Y.}\ \bibnamefont
  {Guan}}, \bibinfo {author} {\bibfnamefont {P.-J.}\ \bibnamefont {Chen}},
  \bibinfo {author} {\bibfnamefont {M.-W.}\ \bibnamefont {Chu}}, \bibinfo
  {author} {\bibfnamefont {R.}~\bibnamefont {Sankar}}, \bibinfo {author}
  {\bibfnamefont {F.}~\bibnamefont {Chou}}, \bibinfo {author} {\bibfnamefont
  {H.-T.}\ \bibnamefont {Jeng}}, \bibinfo {author} {\bibfnamefont {C.-S.}\
  \bibnamefont {Chang}},\ and\ \bibinfo {author} {\bibfnamefont {T.-M.}\
  \bibnamefont {Chuang}},\ }\href
  {https://advances.sciencemag.org/content/2/11/e1600894/tab-article-info}
  {\bibfield  {journal} {\bibinfo  {journal} {Sci. Adv.}\ }\textbf {\bibinfo
  {volume} {2}},\ \bibinfo {pages} {e1600894} (\bibinfo {year}
  {2016})}\BibitemShut {NoStop}%
\bibitem [{\citenamefont {Chen}\ \emph {et~al.}(2016)\citenamefont {Chen},
  \citenamefont {Chang},\ and\ \citenamefont {Jeng}}]{Chen2016}%
  \BibitemOpen
  \bibfield  {author} {\bibinfo {author} {\bibfnamefont {P.-J.}\ \bibnamefont
  {Chen}}, \bibinfo {author} {\bibfnamefont {T.-R.}\ \bibnamefont {Chang}},\
  and\ \bibinfo {author} {\bibfnamefont {H.-T.}\ \bibnamefont {Jeng}},\ }\href
  {https://doi.org/10.1103/PhysRevB.94.165148} {\bibfield  {journal} {\bibinfo
  {journal} {Phys. Rev. B}\ }\textbf {\bibinfo {volume} {94}},\ \bibinfo
  {pages} {165148} (\bibinfo {year} {2016})}\BibitemShut {NoStop}%
\bibitem [{\citenamefont {Chang}\ \emph {et~al.}(2016)\citenamefont {Chang},
  \citenamefont {Chen}, \citenamefont {Bian}, \citenamefont {Huang},
  \citenamefont {Zheng}, \citenamefont {Neupert}, \citenamefont {Sankar},
  \citenamefont {Xu}, \citenamefont {Belopolski}, \citenamefont {Chang},
  \citenamefont {Wang}, \citenamefont {Chou}, \citenamefont {Bansil},
  \citenamefont {Jeng}, \citenamefont {Lin},\ and\ \citenamefont
  {Hasan}}]{Chang2016}%
  \BibitemOpen
  \bibfield  {author} {\bibinfo {author} {\bibfnamefont {T.-R.}\ \bibnamefont
  {Chang}}, \bibinfo {author} {\bibfnamefont {P.-J.}\ \bibnamefont {Chen}},
  \bibinfo {author} {\bibfnamefont {G.}~\bibnamefont {Bian}}, \bibinfo {author}
  {\bibfnamefont {S.-M.}\ \bibnamefont {Huang}}, \bibinfo {author}
  {\bibfnamefont {H.}~\bibnamefont {Zheng}}, \bibinfo {author} {\bibfnamefont
  {T.}~\bibnamefont {Neupert}}, \bibinfo {author} {\bibfnamefont
  {R.}~\bibnamefont {Sankar}}, \bibinfo {author} {\bibfnamefont {S.-Y.}\
  \bibnamefont {Xu}}, \bibinfo {author} {\bibfnamefont {I.}~\bibnamefont
  {Belopolski}}, \bibinfo {author} {\bibfnamefont {G.}~\bibnamefont {Chang}},
  \bibinfo {author} {\bibfnamefont {B.}~\bibnamefont {Wang}}, \bibinfo {author}
  {\bibfnamefont {F.}~\bibnamefont {Chou}}, \bibinfo {author} {\bibfnamefont
  {A.}~\bibnamefont {Bansil}}, \bibinfo {author} {\bibfnamefont {H.-T.}\
  \bibnamefont {Jeng}}, \bibinfo {author} {\bibfnamefont {H.}~\bibnamefont
  {Lin}},\ and\ \bibinfo {author} {\bibfnamefont {M.~Z.}\ \bibnamefont
  {Hasan}},\ }\href {https://doi.org/10.1103/PhysRevB.93.245130} {\bibfield
  {journal} {\bibinfo  {journal} {Phys. Rev. B}\ }\textbf {\bibinfo {volume}
  {93}},\ \bibinfo {pages} {245130} (\bibinfo {year} {2016})}\BibitemShut
  {NoStop}%
\bibitem [{\citenamefont {Wang}\ \emph {et~al.}(2018)\citenamefont {Wang},
  \citenamefont {Liu}, \citenamefont {Zheng}, \citenamefont {Yin},\ and\
  \citenamefont {Wang}}]{Wang2018b}%
  \BibitemOpen
  \bibfield  {author} {\bibinfo {author} {\bibfnamefont {B.-T.}\ \bibnamefont
  {Wang}}, \bibinfo {author} {\bibfnamefont {P.-F.}\ \bibnamefont {Liu}},
  \bibinfo {author} {\bibfnamefont {J.-J.}\ \bibnamefont {Zheng}}, \bibinfo
  {author} {\bibfnamefont {W.}~\bibnamefont {Yin}},\ and\ \bibinfo {author}
  {\bibfnamefont {F.}~\bibnamefont {Wang}},\ }\href
  {https://doi.org/10.1103/PhysRevB.98.014514} {\bibfield  {journal} {\bibinfo
  {journal} {Phys. Rev. B}\ }\textbf {\bibinfo {volume} {98}},\ \bibinfo
  {pages} {014514} (\bibinfo {year} {2018})}\BibitemShut {NoStop}%
\bibitem [{\citenamefont {Chen}\ \emph {et~al.}(2019)\citenamefont {Chen},
  \citenamefont {Wu}, \citenamefont {Jin}, \citenamefont {Li}, \citenamefont
  {Wang}, \citenamefont {Duan}, \citenamefont {Han}, \citenamefont {Li},
  \citenamefont {Long}, \citenamefont {Zhang}, \citenamefont {Chen},\ and\
  \citenamefont {Teng}}]{Chen2019}%
  \BibitemOpen
  \bibfield  {author} {\bibinfo {author} {\bibfnamefont {D.-Y.}\ \bibnamefont
  {Chen}}, \bibinfo {author} {\bibfnamefont {Y.}~\bibnamefont {Wu}}, \bibinfo
  {author} {\bibfnamefont {L.}~\bibnamefont {Jin}}, \bibinfo {author}
  {\bibfnamefont {Y.}~\bibnamefont {Li}}, \bibinfo {author} {\bibfnamefont
  {X.}~\bibnamefont {Wang}}, \bibinfo {author} {\bibfnamefont {J.}~\bibnamefont
  {Duan}}, \bibinfo {author} {\bibfnamefont {J.}~\bibnamefont {Han}}, \bibinfo
  {author} {\bibfnamefont {X.}~\bibnamefont {Li}}, \bibinfo {author}
  {\bibfnamefont {Y.-Z.}\ \bibnamefont {Long}}, \bibinfo {author}
  {\bibfnamefont {X.}~\bibnamefont {Zhang}}, \bibinfo {author} {\bibfnamefont
  {D.}~\bibnamefont {Chen}},\ and\ \bibinfo {author} {\bibfnamefont
  {B.}~\bibnamefont {Teng}},\ }\href
  {https://doi.org/10.1103/PhysRevB.100.064516} {\bibfield  {journal} {\bibinfo
   {journal} {Phys. Rev. B}\ }\textbf {\bibinfo {volume} {100}},\ \bibinfo
  {pages} {064516} (\bibinfo {year} {2019})}\BibitemShut {NoStop}%
\bibitem [{\citenamefont {Feig}\ \emph {et~al.}(2020)\citenamefont {Feig},
  \citenamefont {Baenitz}, \citenamefont {Bobnar}, \citenamefont {L\"uders},
  \citenamefont {Naumann}, \citenamefont {Schnelle}, \citenamefont {Medvediev},
  \citenamefont {Ranjith}, \citenamefont {Hassinger}, \citenamefont {Weigel},
  \citenamefont {Meyer}, \citenamefont {Leithe-Jasper}, \citenamefont
  {Kortus},\ and\ \citenamefont {Gumeniuk}}]{Feig2020}%
  \BibitemOpen
  \bibfield  {author} {\bibinfo {author} {\bibfnamefont {M.}~\bibnamefont
  {Feig}}, \bibinfo {author} {\bibfnamefont {M.}~\bibnamefont {Baenitz}},
  \bibinfo {author} {\bibfnamefont {M.}~\bibnamefont {Bobnar}}, \bibinfo
  {author} {\bibfnamefont {K.}~\bibnamefont {L\"uders}}, \bibinfo {author}
  {\bibfnamefont {M.}~\bibnamefont {Naumann}}, \bibinfo {author} {\bibfnamefont
  {W.}~\bibnamefont {Schnelle}}, \bibinfo {author} {\bibfnamefont
  {S.}~\bibnamefont {Medvediev}}, \bibinfo {author} {\bibfnamefont {K.~M.}\
  \bibnamefont {Ranjith}}, \bibinfo {author} {\bibfnamefont {E.}~\bibnamefont
  {Hassinger}}, \bibinfo {author} {\bibfnamefont {T.}~\bibnamefont {Weigel}},
  \bibinfo {author} {\bibfnamefont {D.~C.}\ \bibnamefont {Meyer}}, \bibinfo
  {author} {\bibfnamefont {A.}~\bibnamefont {Leithe-Jasper}}, \bibinfo {author}
  {\bibfnamefont {J.}~\bibnamefont {Kortus}},\ and\ \bibinfo {author}
  {\bibfnamefont {R.}~\bibnamefont {Gumeniuk}},\ }\href
  {https://doi.org/10.1103/PhysRevB.102.214501} {\bibfield  {journal} {\bibinfo
   {journal} {Phys. Rev. B}\ }\textbf {\bibinfo {volume} {102}},\ \bibinfo
  {pages} {214501} (\bibinfo {year} {2020})}\BibitemShut {NoStop}%
\bibitem [{\citenamefont {Chen}\ \emph {et~al.}(2021)\citenamefont {Chen},
  \citenamefont {Liu}, \citenamefont {Yang}, \citenamefont {Chen},
  \citenamefont {Liu}, \citenamefont {Huang}, \citenamefont {Zhang},
  \citenamefont {Zhang}, \citenamefont {Liu},\ and\ \citenamefont
  {Shen}}]{Chen2021}%
  \BibitemOpen
  \bibfield  {author} {\bibinfo {author} {\bibfnamefont {W.}~\bibnamefont
  {Chen}}, \bibinfo {author} {\bibfnamefont {L.}~\bibnamefont {Liu}}, \bibinfo
  {author} {\bibfnamefont {W.}~\bibnamefont {Yang}}, \bibinfo {author}
  {\bibfnamefont {D.}~\bibnamefont {Chen}}, \bibinfo {author} {\bibfnamefont
  {Z.}~\bibnamefont {Liu}}, \bibinfo {author} {\bibfnamefont {Y.}~\bibnamefont
  {Huang}}, \bibinfo {author} {\bibfnamefont {T.}~\bibnamefont {Zhang}},
  \bibinfo {author} {\bibfnamefont {H.}~\bibnamefont {Zhang}}, \bibinfo
  {author} {\bibfnamefont {Z.}~\bibnamefont {Liu}},\ and\ \bibinfo {author}
  {\bibfnamefont {D.~W.}\ \bibnamefont {Shen}},\ }\href
  {https://doi.org/10.1103/PhysRevB.103.035133} {\bibfield  {journal} {\bibinfo
   {journal} {Phys. Rev. B}\ }\textbf {\bibinfo {volume} {103}},\ \bibinfo
  {pages} {035133} (\bibinfo {year} {2021})}\BibitemShut {NoStop}%
\bibitem [{\citenamefont {Gao}\ \emph {et~al.}(2020)\citenamefont {Gao},
  \citenamefont {Si}, \citenamefont {Luo}, \citenamefont {Yan}, \citenamefont
  {Jiang}, \citenamefont {Wang}, \citenamefont {Xu}, \citenamefont {Xu},
  \citenamefont {Tong}, \citenamefont {Song}, \citenamefont {Zhu},
  \citenamefont {Lu},\ and\ \citenamefont {Sun}}]{Gao2020}%
  \BibitemOpen
  \bibfield  {author} {\bibinfo {author} {\bibfnamefont {J.~J.}\ \bibnamefont
  {Gao}}, \bibinfo {author} {\bibfnamefont {J.~G.}\ \bibnamefont {Si}},
  \bibinfo {author} {\bibfnamefont {X.}~\bibnamefont {Luo}}, \bibinfo {author}
  {\bibfnamefont {J.}~\bibnamefont {Yan}}, \bibinfo {author} {\bibfnamefont
  {Z.~Z.}\ \bibnamefont {Jiang}}, \bibinfo {author} {\bibfnamefont
  {W.}~\bibnamefont {Wang}}, \bibinfo {author} {\bibfnamefont {C.~Q.}\
  \bibnamefont {Xu}}, \bibinfo {author} {\bibfnamefont {X.~F.}\ \bibnamefont
  {Xu}}, \bibinfo {author} {\bibfnamefont {P.}~\bibnamefont {Tong}}, \bibinfo
  {author} {\bibfnamefont {W.~H.}\ \bibnamefont {Song}}, \bibinfo {author}
  {\bibfnamefont {X.~B.}\ \bibnamefont {Zhu}}, \bibinfo {author} {\bibfnamefont
  {W.~J.}\ \bibnamefont {Lu}},\ and\ \bibinfo {author} {\bibfnamefont {Y.~P.}\
  \bibnamefont {Sun}},\ }\href {https://doi.org/10.1021/acs.jpcc.0c00527}
  {\bibfield  {journal} {\bibinfo  {journal} {J. Phys. Chem. C}\ }\textbf
  {\bibinfo {volume} {124}},\ \bibinfo {pages} {6349} (\bibinfo {year}
  {2020})}\BibitemShut {NoStop}%
\bibitem [{\citenamefont {Eppinga}\ and\ \citenamefont
  {Wiegers}(1977)}]{Eppinga1977}%
  \BibitemOpen
  \bibfield  {author} {\bibinfo {author} {\bibfnamefont {R.}~\bibnamefont
  {Eppinga}}\ and\ \bibinfo {author} {\bibfnamefont {G.}~\bibnamefont
  {Wiegers}},\ }\href
  {https://doi.org/https://doi.org/10.1016/0025-5408(77)90033-2} {\bibfield
  {journal} {\bibinfo  {journal} {Mater. Res. Bull.}\ }\textbf {\bibinfo
  {volume} {12}},\ \bibinfo {pages} {1057} (\bibinfo {year}
  {1977})}\BibitemShut {NoStop}%
\bibitem [{\citenamefont {{van der Lee}}\ and\ \citenamefont
  {Wiegers}(1990)}]{Lee1990}%
  \BibitemOpen
  \bibfield  {author} {\bibinfo {author} {\bibfnamefont {A.}~\bibnamefont {{van
  der Lee}}}\ and\ \bibinfo {author} {\bibfnamefont {G.~A.}\ \bibnamefont
  {Wiegers}},\ }\href
  {https://doi.org/https://doi.org/10.1016/0025-5408(90)90008-P} {\bibfield
  {journal} {\bibinfo  {journal} {Mater. Res. Bull.}\ }\textbf {\bibinfo
  {volume} {25}},\ \bibinfo {pages} {1011} (\bibinfo {year}
  {1990})}\BibitemShut {NoStop}%
\bibitem [{\citenamefont {Di~Salvo}\ \emph {et~al.}(1973)\citenamefont
  {Di~Salvo}, \citenamefont {Hull}, \citenamefont {Schwartz}, \citenamefont
  {Voorhoeve},\ and\ \citenamefont {Waszczak}}]{Salvo1973}%
  \BibitemOpen
  \bibfield  {author} {\bibinfo {author} {\bibfnamefont {F.~J.}\ \bibnamefont
  {Di~Salvo}}, \bibinfo {author} {\bibfnamefont {G.~W.}\ \bibnamefont {Hull}},
  \bibinfo {author} {\bibfnamefont {L.~H.}\ \bibnamefont {Schwartz}}, \bibinfo
  {author} {\bibfnamefont {J.~M.}\ \bibnamefont {Voorhoeve}},\ and\ \bibinfo
  {author} {\bibfnamefont {J.~V.}\ \bibnamefont {Waszczak}},\ }\href
  {https://doi.org/10.1063/1.1680277} {\bibfield  {journal} {\bibinfo
  {journal} {J. Chem. Phys.}\ }\textbf {\bibinfo {volume} {59}},\ \bibinfo
  {pages} {1922} (\bibinfo {year} {1973})}\BibitemShut {NoStop}%
\bibitem [{\citenamefont {Eppinga}\ \emph {et~al.}(1981)\citenamefont
  {Eppinga}, \citenamefont {Wiegers},\ and\ \citenamefont
  {Haas}}]{Eppinga1981}%
  \BibitemOpen
  \bibfield  {author} {\bibinfo {author} {\bibfnamefont {R.}~\bibnamefont
  {Eppinga}}, \bibinfo {author} {\bibfnamefont {G.}~\bibnamefont {Wiegers}},\
  and\ \bibinfo {author} {\bibfnamefont {C.}~\bibnamefont {Haas}},\ }\href
  {https://doi.org/https://doi.org/10.1016/0378-4363(81)90240-0} {\bibfield
  {journal} {\bibinfo  {journal} {Physica B+C}\ }\textbf {\bibinfo {volume}
  {105}},\ \bibinfo {pages} {174} (\bibinfo {year} {1981})}\BibitemShut
  {NoStop}%
\bibitem [{\citenamefont {Dijkstra}\ \emph {et~al.}(1989)\citenamefont
  {Dijkstra}, \citenamefont {Broekhuizen}, \citenamefont {van Bruggen},
  \citenamefont {Haas}, \citenamefont {de~Groot},\ and\ \citenamefont {van~der
  Meulen}}]{Dijkstra1989}%
  \BibitemOpen
  \bibfield  {author} {\bibinfo {author} {\bibfnamefont {J.}~\bibnamefont
  {Dijkstra}}, \bibinfo {author} {\bibfnamefont {E.~A.}\ \bibnamefont
  {Broekhuizen}}, \bibinfo {author} {\bibfnamefont {C.~F.}\ \bibnamefont {van
  Bruggen}}, \bibinfo {author} {\bibfnamefont {C.}~\bibnamefont {Haas}},
  \bibinfo {author} {\bibfnamefont {R.~A.}\ \bibnamefont {de~Groot}},\ and\
  \bibinfo {author} {\bibfnamefont {H.~P.}\ \bibnamefont {van~der Meulen}},\
  }\href {https://doi.org/10.1103/PhysRevB.40.12111} {\bibfield  {journal}
  {\bibinfo  {journal} {Phys. Rev. B}\ }\textbf {\bibinfo {volume} {40}},\
  \bibinfo {pages} {12111} (\bibinfo {year} {1989})}\BibitemShut {NoStop}%
\bibitem [{\citenamefont {Eppinga}\ \emph {et~al.}(1976)\citenamefont
  {Eppinga}, \citenamefont {Sawatzky}, \citenamefont {Haas},\ and\
  \citenamefont {van Bruggen}}]{Eppinga1976}%
  \BibitemOpen
  \bibfield  {author} {\bibinfo {author} {\bibfnamefont {R.}~\bibnamefont
  {Eppinga}}, \bibinfo {author} {\bibfnamefont {G.~A.}\ \bibnamefont
  {Sawatzky}}, \bibinfo {author} {\bibfnamefont {C.}~\bibnamefont {Haas}},\
  and\ \bibinfo {author} {\bibfnamefont {C.~F.}\ \bibnamefont {van Bruggen}},\
  }\href {https://doi.org/10.1088/0022-3719/9/17/028} {\bibfield  {journal}
  {\bibinfo  {journal} {J. Phys. C: Solid State Physics}\ }\textbf {\bibinfo
  {volume} {9}},\ \bibinfo {pages} {3371} (\bibinfo {year} {1976})}\BibitemShut
  {NoStop}%
\bibitem [{\citenamefont {Herber}\ and\ \citenamefont
  {Davis}(1975)}]{Herber1975}%
  \BibitemOpen
  \bibfield  {author} {\bibinfo {author} {\bibfnamefont {R.~H.}\ \bibnamefont
  {Herber}}\ and\ \bibinfo {author} {\bibfnamefont {R.~F.}\ \bibnamefont
  {Davis}},\ }\href {https://doi.org/10.1063/1.431763} {\bibfield  {journal}
  {\bibinfo  {journal} {J. Chem. Phys.}\ }\textbf {\bibinfo {volume} {63}},\
  \bibinfo {pages} {3668} (\bibinfo {year} {1975})}\BibitemShut {NoStop}%
\bibitem [{\citenamefont {Herber}\ and\ \citenamefont
  {Davis}(1976)}]{Herber1976}%
  \BibitemOpen
  \bibfield  {author} {\bibinfo {author} {\bibfnamefont {R.~H.}\ \bibnamefont
  {Herber}}\ and\ \bibinfo {author} {\bibfnamefont {R.~F.}\ \bibnamefont
  {Davis}},\ }\href {https://doi.org/10.1063/1.433567} {\bibfield  {journal}
  {\bibinfo  {journal} {J. Chem. Phys.}\ }\textbf {\bibinfo {volume} {65}},\
  \bibinfo {pages} {3773} (\bibinfo {year} {1976})}\BibitemShut {NoStop}%
\bibitem [{\citenamefont {Gentile}\ \emph {et~al.}(1979)\citenamefont
  {Gentile}, \citenamefont {Driscoll},\ and\ \citenamefont
  {Hockman}}]{Gentile1979}%
  \BibitemOpen
  \bibfield  {author} {\bibinfo {author} {\bibfnamefont {P.}~\bibnamefont
  {Gentile}}, \bibinfo {author} {\bibfnamefont {D.}~\bibnamefont {Driscoll}},\
  and\ \bibinfo {author} {\bibfnamefont {A.}~\bibnamefont {Hockman}},\ }\href
  {https://doi.org/https://doi.org/10.1016/S0020-1693(00)93447-9} {\bibfield
  {journal} {\bibinfo  {journal} {Inorg. Chim. Acta}\ }\textbf {\bibinfo
  {volume} {35}},\ \bibinfo {pages} {249} (\bibinfo {year} {1979})}\BibitemShut
  {NoStop}%
\bibitem [{\citenamefont {Herber}\ \emph {et~al.}(1980)\citenamefont {Herber},
  \citenamefont {DiSalvo},\ and\ \citenamefont {Frankel}}]{Herber1980}%
  \BibitemOpen
  \bibfield  {author} {\bibinfo {author} {\bibfnamefont {R.~H.}\ \bibnamefont
  {Herber}}, \bibinfo {author} {\bibfnamefont {F.~J.}\ \bibnamefont
  {DiSalvo}},\ and\ \bibinfo {author} {\bibfnamefont {R.~B.}\ \bibnamefont
  {Frankel}},\ }\href {https://doi.org/10.1021/ic50212a062} {\bibfield
  {journal} {\bibinfo  {journal} {Inorg. Chem.}\ }\textbf {\bibinfo {volume}
  {19}},\ \bibinfo {pages} {3135} (\bibinfo {year} {1980})}\BibitemShut
  {NoStop}%
\bibitem [{\citenamefont {Gossard}\ \emph {et~al.}(1974)\citenamefont
  {Gossard}, \citenamefont {Salvo},\ and\ \citenamefont
  {Yasuoka}}]{Gossard1974}%
  \BibitemOpen
  \bibfield  {author} {\bibinfo {author} {\bibfnamefont {A.~C.}\ \bibnamefont
  {Gossard}}, \bibinfo {author} {\bibfnamefont {F.~J.~d.}\ \bibnamefont
  {Salvo}},\ and\ \bibinfo {author} {\bibfnamefont {H.}~\bibnamefont
  {Yasuoka}},\ }\href {https://doi.org/10.1103/PhysRevB.9.3965} {\bibfield
  {journal} {\bibinfo  {journal} {Phys. Rev. B}\ }\textbf {\bibinfo {volume}
  {9}},\ \bibinfo {pages} {3965} (\bibinfo {year} {1974})}\BibitemShut
  {NoStop}%
\bibitem [{\citenamefont {Guo}\ and\ \citenamefont {Liang}(1987)}]{Guo1987}%
  \BibitemOpen
  \bibfield  {author} {\bibinfo {author} {\bibfnamefont {G.~Y.}\ \bibnamefont
  {Guo}}\ and\ \bibinfo {author} {\bibfnamefont {W.~Y.}\ \bibnamefont
  {Liang}},\ }\href {https://doi.org/10.1088/0022-3719/20/27/011} {\bibfield
  {journal} {\bibinfo  {journal} {J. Phys. C: Solid State Phys.}\ }\textbf
  {\bibinfo {volume} {20}},\ \bibinfo {pages} {4315} (\bibinfo {year}
  {1987})}\BibitemShut {NoStop}%
\bibitem [{\citenamefont {Blaha}(1991)}]{Blaha1991}%
  \BibitemOpen
  \bibfield  {author} {\bibinfo {author} {\bibfnamefont {P.}~\bibnamefont
  {Blaha}},\ }\href {https://doi.org/10.1088/0953-8984/3/47/011} {\bibfield
  {journal} {\bibinfo  {journal} {J. Phys.: Condens. Matter}\ }\textbf
  {\bibinfo {volume} {3}},\ \bibinfo {pages} {9381} (\bibinfo {year}
  {1991})}\BibitemShut {NoStop}%
\bibitem [{Top()}]{Topas}%
  \BibitemOpen
  \href@noop {} {\bibinfo {title} {{Bruker AXS, TOPAS 4.2, Karlsruhe, Germany,
  2009}}}\BibitemShut {NoStop}%
\bibitem [{\citenamefont {Kresse}\ and\ \citenamefont
  {Hafner}(1993)}]{Kresse1993}%
  \BibitemOpen
  \bibfield  {author} {\bibinfo {author} {\bibfnamefont {G.}~\bibnamefont
  {Kresse}}\ and\ \bibinfo {author} {\bibfnamefont {J.}~\bibnamefont
  {Hafner}},\ }\href {https://doi.org/10.1103/PhysRevB.47.558} {\bibfield
  {journal} {\bibinfo  {journal} {Phys. Rev. B}\ }\textbf {\bibinfo {volume}
  {47}},\ \bibinfo {pages} {558} (\bibinfo {year} {1993})}\BibitemShut
  {NoStop}%
\bibitem [{\citenamefont {Kresse}\ and\ \citenamefont
  {Hafner}(1994{\natexlab{a}})}]{Kresse1994}%
  \BibitemOpen
  \bibfield  {author} {\bibinfo {author} {\bibfnamefont {G.}~\bibnamefont
  {Kresse}}\ and\ \bibinfo {author} {\bibfnamefont {J.}~\bibnamefont
  {Hafner}},\ }\href {https://doi.org/10.1103/PhysRevB.49.14251} {\bibfield
  {journal} {\bibinfo  {journal} {Phys. Rev. B}\ }\textbf {\bibinfo {volume}
  {49}},\ \bibinfo {pages} {14251} (\bibinfo {year}
  {1994}{\natexlab{a}})}\BibitemShut {NoStop}%
\bibitem [{\citenamefont {Kresse}\ and\ \citenamefont
  {Furthm\"{u}ller}(1996)}]{Kresse1996}%
  \BibitemOpen
  \bibfield  {author} {\bibinfo {author} {\bibfnamefont {G.}~\bibnamefont
  {Kresse}}\ and\ \bibinfo {author} {\bibfnamefont {J.}~\bibnamefont
  {Furthm\"{u}ller}},\ }\href
  {https://doi.org/https://doi.org/10.1016/0927-0256(96)00008-0} {\bibfield
  {journal} {\bibinfo  {journal} {Comput. Mater. Sci.}\ }\textbf {\bibinfo
  {volume} {6}},\ \bibinfo {pages} {15 } (\bibinfo {year} {1996})}\BibitemShut
  {NoStop}%
\bibitem [{\citenamefont {Kresse}\ and\ \citenamefont
  {Furthm\"uller}(1996)}]{Kresse1996b}%
  \BibitemOpen
  \bibfield  {author} {\bibinfo {author} {\bibfnamefont {G.}~\bibnamefont
  {Kresse}}\ and\ \bibinfo {author} {\bibfnamefont {J.}~\bibnamefont
  {Furthm\"uller}},\ }\href {https://doi.org/10.1103/PhysRevB.54.11169}
  {\bibfield  {journal} {\bibinfo  {journal} {Phys. Rev. B}\ }\textbf {\bibinfo
  {volume} {54}},\ \bibinfo {pages} {11169} (\bibinfo {year}
  {1996})}\BibitemShut {NoStop}%
\bibitem [{\citenamefont {Kresse}\ and\ \citenamefont
  {Hafner}(1994{\natexlab{b}})}]{Kresse1994b}%
  \BibitemOpen
  \bibfield  {author} {\bibinfo {author} {\bibfnamefont {G.}~\bibnamefont
  {Kresse}}\ and\ \bibinfo {author} {\bibfnamefont {J.}~\bibnamefont
  {Hafner}},\ }\href {https://doi.org/10.1088/0953-8984/6/40/015} {\bibfield
  {journal} {\bibinfo  {journal} {J. Phys. Condens. Matter}\ }\textbf {\bibinfo
  {volume} {6}},\ \bibinfo {pages} {8245} (\bibinfo {year}
  {1994}{\natexlab{b}})}\BibitemShut {NoStop}%
\bibitem [{\citenamefont {Kresse}\ and\ \citenamefont
  {Joubert}(1999)}]{Kresse1999}%
  \BibitemOpen
  \bibfield  {author} {\bibinfo {author} {\bibfnamefont {G.}~\bibnamefont
  {Kresse}}\ and\ \bibinfo {author} {\bibfnamefont {D.}~\bibnamefont
  {Joubert}},\ }\href {https://doi.org/10.1103/PhysRevB.59.1758} {\bibfield
  {journal} {\bibinfo  {journal} {Phys. Rev. B}\ }\textbf {\bibinfo {volume}
  {59}},\ \bibinfo {pages} {1758} (\bibinfo {year} {1999})}\BibitemShut
  {NoStop}%
\bibitem [{\citenamefont {Grimme}\ \emph {et~al.}(2010)\citenamefont {Grimme},
  \citenamefont {Antony}, \citenamefont {Ehrlich},\ and\ \citenamefont
  {Krieg}}]{Grimme2010}%
  \BibitemOpen
  \bibfield  {author} {\bibinfo {author} {\bibfnamefont {S.}~\bibnamefont
  {Grimme}}, \bibinfo {author} {\bibfnamefont {J.}~\bibnamefont {Antony}},
  \bibinfo {author} {\bibfnamefont {S.}~\bibnamefont {Ehrlich}},\ and\ \bibinfo
  {author} {\bibfnamefont {H.}~\bibnamefont {Krieg}},\ }\href
  {https://doi.org/10.1063/1.3382344} {\bibfield  {journal} {\bibinfo
  {journal} {J. Chem. Phys.}\ }\textbf {\bibinfo {volume} {132}},\ \bibinfo
  {pages} {154104} (\bibinfo {year} {2010})}\BibitemShut {NoStop}%
\bibitem [{\citenamefont {Grimme}\ \emph {et~al.}(2011)\citenamefont {Grimme},
  \citenamefont {Ehrlich},\ and\ \citenamefont {Goerigk}}]{Grimme2011}%
  \BibitemOpen
  \bibfield  {author} {\bibinfo {author} {\bibfnamefont {S.}~\bibnamefont
  {Grimme}}, \bibinfo {author} {\bibfnamefont {S.}~\bibnamefont {Ehrlich}},\
  and\ \bibinfo {author} {\bibfnamefont {L.}~\bibnamefont {Goerigk}},\ }\href
  {https://doi.org/https://doi.org/10.1002/jcc.21759} {\bibfield  {journal}
  {\bibinfo  {journal} {J. Comput. Chem.}\ }\textbf {\bibinfo {volume} {32}},\
  \bibinfo {pages} {1456} (\bibinfo {year} {2011})}\BibitemShut {NoStop}%
\bibitem [{\citenamefont {Meetsma}\ \emph {et~al.}(1990)\citenamefont
  {Meetsma}, \citenamefont {Wiegers}, \citenamefont {Haange},\ and\
  \citenamefont {de~Boer}}]{Meetsma1990}%
  \BibitemOpen
  \bibfield  {author} {\bibinfo {author} {\bibfnamefont {A.}~\bibnamefont
  {Meetsma}}, \bibinfo {author} {\bibfnamefont {G.~A.}\ \bibnamefont
  {Wiegers}}, \bibinfo {author} {\bibfnamefont {R.~J.}\ \bibnamefont
  {Haange}},\ and\ \bibinfo {author} {\bibfnamefont {J.~L.}\ \bibnamefont
  {de~Boer}},\ }\href {https://doi.org/10.1107/S0108270190000014} {\bibfield
  {journal} {\bibinfo  {journal} {Acta Cryst.}\ }\textbf {\bibinfo {volume} {C
  46}},\ \bibinfo {pages} {1598} (\bibinfo {year} {1990})}\BibitemShut
  {NoStop}%
\bibitem [{\citenamefont {Gurevich}(2003)}]{Gurevich2003}%
  \BibitemOpen
  \bibfield  {author} {\bibinfo {author} {\bibfnamefont {A.}~\bibnamefont
  {Gurevich}},\ }\href {https://doi.org/10.1103/PhysRevB.67.184515} {\bibfield
  {journal} {\bibinfo  {journal} {Phys. Rev. B}\ }\textbf {\bibinfo {volume}
  {67}},\ \bibinfo {pages} {184515} (\bibinfo {year} {2003})}\BibitemShut
  {NoStop}%
\bibitem [{\citenamefont {Lei}\ \emph {et~al.}(2012)\citenamefont {Lei},
  \citenamefont {Graf}, \citenamefont {Hu}, \citenamefont {Ryu}, \citenamefont
  {Choi}, \citenamefont {Tozer},\ and\ \citenamefont {Petrovic}}]{Lei2012}%
  \BibitemOpen
  \bibfield  {author} {\bibinfo {author} {\bibfnamefont {H.}~\bibnamefont
  {Lei}}, \bibinfo {author} {\bibfnamefont {D.}~\bibnamefont {Graf}}, \bibinfo
  {author} {\bibfnamefont {R.}~\bibnamefont {Hu}}, \bibinfo {author}
  {\bibfnamefont {H.}~\bibnamefont {Ryu}}, \bibinfo {author} {\bibfnamefont
  {E.~S.}\ \bibnamefont {Choi}}, \bibinfo {author} {\bibfnamefont {S.~W.}\
  \bibnamefont {Tozer}},\ and\ \bibinfo {author} {\bibfnamefont
  {C.}~\bibnamefont {Petrovic}},\ }\href
  {https://doi.org/10.1103/PhysRevB.85.094515} {\bibfield  {journal} {\bibinfo
  {journal} {Phys. Rev. B}\ }\textbf {\bibinfo {volume} {85}},\ \bibinfo
  {pages} {094515} (\bibinfo {year} {2012})}\BibitemShut {NoStop}%
\bibitem [{\citenamefont {Karki}\ \emph {et~al.}(2011)\citenamefont {Karki},
  \citenamefont {Xiong}, \citenamefont {Haldolaarachchige}, \citenamefont
  {Stadler}, \citenamefont {Vekhter}, \citenamefont {Adams}, \citenamefont
  {Young}, \citenamefont {Phelan},\ and\ \citenamefont {Chan}}]{Karki2011}%
  \BibitemOpen
  \bibfield  {author} {\bibinfo {author} {\bibfnamefont {A.~B.}\ \bibnamefont
  {Karki}}, \bibinfo {author} {\bibfnamefont {Y.~M.}\ \bibnamefont {Xiong}},
  \bibinfo {author} {\bibfnamefont {N.}~\bibnamefont {Haldolaarachchige}},
  \bibinfo {author} {\bibfnamefont {S.}~\bibnamefont {Stadler}}, \bibinfo
  {author} {\bibfnamefont {I.}~\bibnamefont {Vekhter}}, \bibinfo {author}
  {\bibfnamefont {P.~W.}\ \bibnamefont {Adams}}, \bibinfo {author}
  {\bibfnamefont {D.~P.}\ \bibnamefont {Young}}, \bibinfo {author}
  {\bibfnamefont {W.~A.}\ \bibnamefont {Phelan}},\ and\ \bibinfo {author}
  {\bibfnamefont {J.~Y.}\ \bibnamefont {Chan}},\ }\href
  {https://doi.org/10.1103/PhysRevB.83.144525} {\bibfield  {journal} {\bibinfo
  {journal} {Phys. Rev. B}\ }\textbf {\bibinfo {volume} {83}},\ \bibinfo
  {pages} {144525} (\bibinfo {year} {2011})}\BibitemShut {NoStop}%
\bibitem [{\citenamefont {Takano}\ \emph {et~al.}(2001)\citenamefont {Takano},
  \citenamefont {Takeya}, \citenamefont {Fujii}, \citenamefont {Kumakura},
  \citenamefont {Hatano}, \citenamefont {Togano}, \citenamefont {Kito},\ and\
  \citenamefont {Ihara}}]{Takano2001}%
  \BibitemOpen
  \bibfield  {author} {\bibinfo {author} {\bibfnamefont {Y.}~\bibnamefont
  {Takano}}, \bibinfo {author} {\bibfnamefont {H.}~\bibnamefont {Takeya}},
  \bibinfo {author} {\bibfnamefont {H.}~\bibnamefont {Fujii}}, \bibinfo
  {author} {\bibfnamefont {H.}~\bibnamefont {Kumakura}}, \bibinfo {author}
  {\bibfnamefont {T.}~\bibnamefont {Hatano}}, \bibinfo {author} {\bibfnamefont
  {K.}~\bibnamefont {Togano}}, \bibinfo {author} {\bibfnamefont
  {H.}~\bibnamefont {Kito}},\ and\ \bibinfo {author} {\bibfnamefont
  {H.}~\bibnamefont {Ihara}},\ }\href {https://doi.org/10.1063/1.1371239}
  {\bibfield  {journal} {\bibinfo  {journal} {Appl. Phys. Lett.}\ }\textbf
  {\bibinfo {volume} {78}},\ \bibinfo {pages} {2914} (\bibinfo {year}
  {2001})}\BibitemShut {NoStop}%
\bibitem [{\citenamefont {Naskar}\ \emph {et~al.}(2022)\citenamefont {Naskar},
  \citenamefont {Ash}, \citenamefont {Panda}, \citenamefont {Vishwakarma},
  \citenamefont {Mani}, \citenamefont {Sundaresan},\ and\ \citenamefont
  {Ganguli}}]{Naskar2022}%
  \BibitemOpen
  \bibfield  {author} {\bibinfo {author} {\bibfnamefont {M.}~\bibnamefont
  {Naskar}}, \bibinfo {author} {\bibfnamefont {S.}~\bibnamefont {Ash}},
  \bibinfo {author} {\bibfnamefont {D.~P.}\ \bibnamefont {Panda}}, \bibinfo
  {author} {\bibfnamefont {C.~K.}\ \bibnamefont {Vishwakarma}}, \bibinfo
  {author} {\bibfnamefont {B.~K.}\ \bibnamefont {Mani}}, \bibinfo {author}
  {\bibfnamefont {A.}~\bibnamefont {Sundaresan}},\ and\ \bibinfo {author}
  {\bibfnamefont {A.~K.}\ \bibnamefont {Ganguli}},\ }\href
  {https://doi.org/10.1103/PhysRevB.105.014513} {\bibfield  {journal} {\bibinfo
   {journal} {Phys. Rev. B}\ }\textbf {\bibinfo {volume} {105}},\ \bibinfo
  {pages} {014513} (\bibinfo {year} {2022})}\BibitemShut {NoStop}%
\bibitem [{\citenamefont {{Lan}}\ \emph {et~al.}(2001)\citenamefont {{Lan}},
  \citenamefont {{Chang}}, \citenamefont {{Lu}}, \citenamefont {{Lee}},
  \citenamefont {{Shih}},\ and\ \citenamefont {{Jeng}}}]{Lan2001}%
  \BibitemOpen
  \bibfield  {author} {\bibinfo {author} {\bibfnamefont {M.~D.}\ \bibnamefont
  {{Lan}}}, \bibinfo {author} {\bibfnamefont {J.~C.}\ \bibnamefont {{Chang}}},
  \bibinfo {author} {\bibfnamefont {K.~T.}\ \bibnamefont {{Lu}}}, \bibinfo
  {author} {\bibfnamefont {C.~Y.}\ \bibnamefont {{Lee}}}, \bibinfo {author}
  {\bibfnamefont {H.~Y.}\ \bibnamefont {{Shih}}},\ and\ \bibinfo {author}
  {\bibfnamefont {G.~Y.}\ \bibnamefont {{Jeng}}},\ }\href
  {https://doi.org/10.1109/77.919845} {\bibfield  {journal} {\bibinfo
  {journal} {IEEE Transactions on Applied Superconductivity}\ }\textbf
  {\bibinfo {volume} {11}},\ \bibinfo {pages} {3607} (\bibinfo {year}
  {2001})}\BibitemShut {NoStop}%
\bibitem [{\citenamefont {Rathnayaka}\ \emph {et~al.}(1997)\citenamefont
  {Rathnayaka}, \citenamefont {Bhatnagar}, \citenamefont {Parasiris},
  \citenamefont {Naugle}, \citenamefont {Canfield},\ and\ \citenamefont
  {Cho}}]{Rathnayaka1997}%
  \BibitemOpen
  \bibfield  {author} {\bibinfo {author} {\bibfnamefont {K.~D.~D.}\
  \bibnamefont {Rathnayaka}}, \bibinfo {author} {\bibfnamefont {A.~K.}\
  \bibnamefont {Bhatnagar}}, \bibinfo {author} {\bibfnamefont {A.}~\bibnamefont
  {Parasiris}}, \bibinfo {author} {\bibfnamefont {D.~G.}\ \bibnamefont
  {Naugle}}, \bibinfo {author} {\bibfnamefont {P.~C.}\ \bibnamefont
  {Canfield}},\ and\ \bibinfo {author} {\bibfnamefont {B.~K.}\ \bibnamefont
  {Cho}},\ }\href {https://doi.org/10.1103/PhysRevB.55.8506} {\bibfield
  {journal} {\bibinfo  {journal} {Phys. Rev. B}\ }\textbf {\bibinfo {volume}
  {55}},\ \bibinfo {pages} {8506} (\bibinfo {year} {1997})}\BibitemShut
  {NoStop}%
\bibitem [{\citenamefont {Micnas}\ \emph {et~al.}(1990)\citenamefont {Micnas},
  \citenamefont {Ranninger},\ and\ \citenamefont {Robaszkiewicz}}]{Micnas1990}%
  \BibitemOpen
  \bibfield  {author} {\bibinfo {author} {\bibfnamefont {R.}~\bibnamefont
  {Micnas}}, \bibinfo {author} {\bibfnamefont {J.}~\bibnamefont {Ranninger}},\
  and\ \bibinfo {author} {\bibfnamefont {S.}~\bibnamefont {Robaszkiewicz}},\
  }\href {https://doi.org/10.1103/RevModPhys.62.113} {\bibfield  {journal}
  {\bibinfo  {journal} {Rev. Mod. Phys.}\ }\textbf {\bibinfo {volume} {62}},\
  \bibinfo {pages} {113} (\bibinfo {year} {1990})}\BibitemShut {NoStop}%
\bibitem [{\citenamefont {Alexandrov}\ \emph {et~al.}(1986)\citenamefont
  {Alexandrov}, \citenamefont {Ranninger},\ and\ \citenamefont
  {Robaszkiewicz}}]{Alexandrov1986}%
  \BibitemOpen
  \bibfield  {author} {\bibinfo {author} {\bibfnamefont {A.~S.}\ \bibnamefont
  {Alexandrov}}, \bibinfo {author} {\bibfnamefont {J.}~\bibnamefont
  {Ranninger}},\ and\ \bibinfo {author} {\bibfnamefont {S.}~\bibnamefont
  {Robaszkiewicz}},\ }\href {https://doi.org/10.1103/PhysRevB.33.4526}
  {\bibfield  {journal} {\bibinfo  {journal} {Phys. Rev. B}\ }\textbf {\bibinfo
  {volume} {33}},\ \bibinfo {pages} {4526} (\bibinfo {year}
  {1986})}\BibitemShut {NoStop}%
\bibitem [{\citenamefont {Alexandrov}(1993)}]{Alexandrov1993}%
  \BibitemOpen
  \bibfield  {author} {\bibinfo {author} {\bibfnamefont {A.~S.}\ \bibnamefont
  {Alexandrov}},\ }\href {https://doi.org/10.1103/PhysRevB.48.10571} {\bibfield
   {journal} {\bibinfo  {journal} {Phys. Rev. B}\ }\textbf {\bibinfo {volume}
  {48}},\ \bibinfo {pages} {10571} (\bibinfo {year} {1993})}\BibitemShut
  {NoStop}%
\bibitem [{\citenamefont {Alexandrov}(2004)}]{Alexandrov2004}%
  \BibitemOpen
  \bibfield  {author} {\bibinfo {author} {\bibfnamefont {A.}~\bibnamefont
  {Alexandrov}},\ }\href
  {https://doi.org/https://doi.org/10.1016/j.physc.2003.10.038} {\bibfield
  {journal} {\bibinfo  {journal} {Physica C: Superconductivity}\ }\textbf
  {\bibinfo {volume} {404}},\ \bibinfo {pages} {22} (\bibinfo {year}
  {2004})}\BibitemShut {NoStop}%
\bibitem [{\citenamefont {Werthamer}\ \emph {et~al.}(1966)\citenamefont
  {Werthamer}, \citenamefont {Helfand},\ and\ \citenamefont
  {Hohenberg}}]{Werthamer1966}%
  \BibitemOpen
  \bibfield  {author} {\bibinfo {author} {\bibfnamefont {N.~R.}\ \bibnamefont
  {Werthamer}}, \bibinfo {author} {\bibfnamefont {E.}~\bibnamefont {Helfand}},\
  and\ \bibinfo {author} {\bibfnamefont {P.~C.}\ \bibnamefont {Hohenberg}},\
  }\href {https://doi.org/10.1103/PhysRev.147.295} {\bibfield  {journal}
  {\bibinfo  {journal} {Phys. Rev.}\ }\textbf {\bibinfo {volume} {147}},\
  \bibinfo {pages} {295} (\bibinfo {year} {1966})}\BibitemShut {NoStop}%
\bibitem [{\citenamefont {Peets}\ \emph {et~al.}(2011)\citenamefont {Peets},
  \citenamefont {Eguchi}, \citenamefont {Kriener}, \citenamefont {Harada},
  \citenamefont {Shamsuzzamen}, \citenamefont {Inada}, \citenamefont {Zheng},\
  and\ \citenamefont {Maeno}}]{Peets2011}%
  \BibitemOpen
  \bibfield  {author} {\bibinfo {author} {\bibfnamefont {D.~C.}\ \bibnamefont
  {Peets}}, \bibinfo {author} {\bibfnamefont {G.}~\bibnamefont {Eguchi}},
  \bibinfo {author} {\bibfnamefont {M.}~\bibnamefont {Kriener}}, \bibinfo
  {author} {\bibfnamefont {S.}~\bibnamefont {Harada}}, \bibinfo {author}
  {\bibfnamefont {S.~M.}\ \bibnamefont {Shamsuzzamen}}, \bibinfo {author}
  {\bibfnamefont {Y.}~\bibnamefont {Inada}}, \bibinfo {author} {\bibfnamefont
  {G.-Q.}\ \bibnamefont {Zheng}},\ and\ \bibinfo {author} {\bibfnamefont
  {Y.}~\bibnamefont {Maeno}},\ }\href
  {https://doi.org/10.1103/PhysRevB.84.054521} {\bibfield  {journal} {\bibinfo
  {journal} {Phys. Rev. B}\ }\textbf {\bibinfo {volume} {84}},\ \bibinfo
  {pages} {054521} (\bibinfo {year} {2011})}\BibitemShut {NoStop}%
\bibitem [{\citenamefont {Clogston}(1962)}]{Clogston1962}%
  \BibitemOpen
  \bibfield  {author} {\bibinfo {author} {\bibfnamefont {A.~M.}\ \bibnamefont
  {Clogston}},\ }\href {https://doi.org/10.1103/PhysRevLett.9.266} {\bibfield
  {journal} {\bibinfo  {journal} {Phys. Rev. Lett.}\ }\textbf {\bibinfo
  {volume} {9}},\ \bibinfo {pages} {266} (\bibinfo {year} {1962})}\BibitemShut
  {NoStop}%
\bibitem [{\citenamefont {Chandrasekhar}(1962)}]{Chandrasekhar1962}%
  \BibitemOpen
  \bibfield  {author} {\bibinfo {author} {\bibfnamefont {B.~S.}\ \bibnamefont
  {Chandrasekhar}},\ }\href {https://doi.org/10.1063/1.1777362} {\bibfield
  {journal} {\bibinfo  {journal} {Appl. Phys. Lett.}\ }\textbf {\bibinfo
  {volume} {1}},\ \bibinfo {pages} {7} (\bibinfo {year} {1962})}\BibitemShut
  {NoStop}%
\bibitem [{\citenamefont {Maki}(1966)}]{Maki1966}%
  \BibitemOpen
  \bibfield  {author} {\bibinfo {author} {\bibfnamefont {K.}~\bibnamefont
  {Maki}},\ }\href {https://doi.org/10.1103/PhysRev.148.362} {\bibfield
  {journal} {\bibinfo  {journal} {Phys. Rev.}\ }\textbf {\bibinfo {volume}
  {148}},\ \bibinfo {pages} {362} (\bibinfo {year} {1966})}\BibitemShut
  {NoStop}%
\bibitem [{\citenamefont {McMillan}(1968)}]{Mcmillan1968}%
  \BibitemOpen
  \bibfield  {author} {\bibinfo {author} {\bibfnamefont {W.~L.}\ \bibnamefont
  {McMillan}},\ }\href {https://doi.org/10.1103/PhysRev.167.331} {\bibfield
  {journal} {\bibinfo  {journal} {Phys. Rev.}\ }\textbf {\bibinfo {volume}
  {167}},\ \bibinfo {pages} {331} (\bibinfo {year} {1968})}\BibitemShut
  {NoStop}%
\bibitem [{\citenamefont {Kadowaki}\ and\ \citenamefont
  {Woods}(1986)}]{Kadowaki1986}%
  \BibitemOpen
  \bibfield  {author} {\bibinfo {author} {\bibfnamefont {K.}~\bibnamefont
  {Kadowaki}}\ and\ \bibinfo {author} {\bibfnamefont {S.}~\bibnamefont
  {Woods}},\ }\href
  {https://doi.org/https://doi.org/10.1016/0038-1098(86)90785-4} {\bibfield
  {journal} {\bibinfo  {journal} {Solid State Commun.}\ }\textbf {\bibinfo
  {volume} {58}},\ \bibinfo {pages} {507} (\bibinfo {year} {1986})}\BibitemShut
  {NoStop}%
\bibitem [{\citenamefont {Jacko}\ \emph {et~al.}(2009)\citenamefont {Jacko},
  \citenamefont {Fj{\ae}restad},\ and\ \citenamefont {Powell}}]{Jacko2009}%
  \BibitemOpen
  \bibfield  {author} {\bibinfo {author} {\bibfnamefont {A.~C.}\ \bibnamefont
  {Jacko}}, \bibinfo {author} {\bibfnamefont {J.~O.}\ \bibnamefont
  {Fj{\ae}restad}},\ and\ \bibinfo {author} {\bibfnamefont {B.~J.}\
  \bibnamefont {Powell}},\ }\href
  {https://doi.org/https://doi.org/10.1038/nphys1249} {\bibfield  {journal}
  {\bibinfo  {journal} {Nat. Phys.}\ }\textbf {\bibinfo {volume} {5}},\
  \bibinfo {pages} {422} (\bibinfo {year} {2009})}\BibitemShut {NoStop}%
\bibitem [{\citenamefont {Singh}\ \emph {et~al.}(2022)\citenamefont {Singh},
  \citenamefont {Saha}, \citenamefont {Nagpal},\ and\ \citenamefont
  {Patnaik}}]{Singh2022}%
  \BibitemOpen
  \bibfield  {author} {\bibinfo {author} {\bibfnamefont {M.}~\bibnamefont
  {Singh}}, \bibinfo {author} {\bibfnamefont {P.}~\bibnamefont {Saha}},
  \bibinfo {author} {\bibfnamefont {V.}~\bibnamefont {Nagpal}},\ and\ \bibinfo
  {author} {\bibfnamefont {S.}~\bibnamefont {Patnaik}},\ }\href
  {https://arxiv.org/abs/2202.10711} {\bibfield  {journal} {\bibinfo  {journal}
  {arXiv:2202.10711[cond-mat.supr-con]}\ } (\bibinfo {year}
  {2022})}\BibitemShut {NoStop}%
\end{thebibliography}

%

\end{document}